\documentclass[]{agujournal2019}
\usepackage{url} 
\usepackage{lineno}
\usepackage{soul}
\usepackage{amssymb}
\usepackage{amsmath}
\linenumbers

\newcommand\Rey{\mbox{\textit{Re}}}  
\newcommand\Pran{\mbox{\textit{Pr}}} 

\newcommand\eg{e.g.\ }
\newcommand\ie{i.e.\ }
\newcommand{\pl}{\left(}
\newcommand{\pr}{\right)}
\newcommand{\n}{\nabla}
\newcommand{\vn}{\boldsymbol{\nabla}}
\newcommand{\vect}[1]{\boldsymbol{#1}}
\newcommand{\vel}{\mathbf{u}}

\newcommand{\zw}{\overline{u}_{\phi}}
\newcommand{\pd}[2]{\frac{\partial #1}{\partial #2}}
\newcommand{\Tpert}{\Theta}
\newcommand{\Cpert}{\xi}

\newcommand{\Ek}{\mbox{\textit{Ek}}}
\newcommand{\Ra}{\mbox{\textit{Ra}}}
\newcommand{\Ro}{\mbox{\textit{Ro}}}
\newcommand{\Nu}{\mbox{\textit{Nu}}}
\newcommand{\Sc}{\mbox{\textit{Sc}}}
\newcommand{\Le}{\mbox{\textit{Le}}}
\newcommand{\Ta}{\mbox{\textit{Ta}}}
\newcommand{\asp}{\chi}
\newcommand{\Rm}{\mbox{\textit{Rm}}}
\newcommand{\Pm}{\mbox{\textit{Pm}}}

\draftfalse
\nolinenumbers

\journalname{JGR: Planets}

\begin{document}

\title{Fingering convection in the stably-stratified layers of planetary cores}

\authors{C\'eline Guervilly\affil{1}}

\affiliation{1}{School of Mathematics, Statistics and Physics, Newcastle University, Newcastle Upon Tyne NE1 7RU, United Kingdom}

\correspondingauthor{C\'eline Guervilly}{celine.guervilly@ncl.ac.uk}

\begin{keypoints}
\item Fingering convection in a Mercury-like stable layer produces thin sheet-like radial flows that are elongated in the meridional direction.
\item The radial flows are always laminar and their azimuthal length is expected to be about 1m in planetary cores.
\item Strong zonal flows form as a by-product of the latitudinal variations of the composition transport.
\end{keypoints}

\begin{abstract}
Stably-stratified layers may be present at the top of the electrically-conducting fluid layers of many planets either because the temperature gradient is locally subadiabatic or because a stable composition gradient is maintained by the segregation of chemical elements. Here we study the double-diffusive processes taking place in such a stable layer, considering the case of Mercury's core where the temperature gradient is stable but the composition gradient is unstable over a 800km-thick layer. The large difference in the molecular diffusivities leads to the development of buoyancy-driven instabilities that drive radial flows known as fingering convection. We model fingering convection using hydrodynamical simulations in a rotating spherical shell and varying the rotation rate and the stratification strength. For small Rayleigh numbers (\ie weak background temperature and composition gradients), fingering convection takes the form of columnar flows aligned with the rotation axis and with an azimuthal size comparable with the layer thickness. For larger Rayleigh numbers, the flows retain a columnar structure but the azimuthal size is drastically reduced leading to thin sheet-like structures that are elongated in the meridional direction. The azimuthal size decreases when the thermal stratification increases, following closely the scaling law expected from the linear planar theory \cite{Stern1960}. We find that the radial flows always remain laminar with local Reynolds number of order $1-10$. Equatorially-symmetric zonal flows form due to latitudinal variations of the axisymmetric composition. The zonal velocity exceeds the non-axisymmetric velocities at the largest Rayleigh numbers. We discuss plausible implications for planetary magnetic fields.
\end{abstract}

\section*{Plain Language Summary}
Convection occurs in planetary interiors due to local changes of density, which can be produced by changes of temperature or chemical composition.  In particular, convection occurs in the electrically-conducting fluid layers located deep inside planets and is at the origin of the generation of planetary magnetic fields. However, for many planets, the upper part of this electrically-conducting region might not be subject to standard convection because the gradients of either temperature or chemical composition produce a further increase of the density with depth, leading to the formation of a stable layer. In some cases, the gradients of these two components act in opposition. Such might be the case of the upper part of Mercury's core, where the stable layer is maintained by the thermal gradient, but the compositional gradient is unstable. This situation is prone to fingering convection, where fluid instabilities release the potential energy associated with the compositional gradient. Here we show that fingering convection consists of sheet-like flows with a narrow 
longitudinal size of approximately 1m in Mercury-like conditions. Strong zonal (\ie east/westward) flows also form. The presence of fingering convection in stable layers might have important consequences for the magnetic fields observed at the planet’s surface.

\section{Introduction}

Planetary magnetic fields are maintained by dynamo action, a process that converts kinetic energy into magnetic energy. 
This process is thought to be mainly driven by convection and occurs in the fluid core \cite{Landeau2022}. 
Here by ``core" we mean the electrically-conducting fluid layer of a planet,
\ie the liquid iron core of terrestrial planets, the metallic hydrogen layer of gas giants and the ice layer of ice giants.
In some planets, the density gradient in the upper region of the core might be stable to standard overturning convection (\ie lighter fluid overlies heavier fluid in this region). 
In this configuration, a stably-stratified layer would therefore surround the convective dynamo region, and this would have profound consequences for the magnetic field observed at the surface
\cite{Christensen2006}.
Indeed, stable density gradients are often thought to damp vertical motions,
so the magnetic field diffusing through the stable top layer from deeper down would be subject to a filtering 
due to electromagnetic skin effects and differential rotation, which damp rapidly-fluctuating and non-axisymmetric fields respectively \cite{Stevenson1982}.
This ``stable top layer scenario" is often used to explain the unusual surface magnetic fields of Mercury and Saturn, which are extremely axisymmetric
\cite{Christensen2008,Stanley2010,Yan2021}, and of Ganymede, which has an anomalously low quadrupole moment \cite{Christensen2015}. 
This scenario has also been applied in the context of the Earth's core, where the presence of a stable top layer is debated, 
and the geomagnetic observations are used to provide constraints on the thickness and stratification of the stable layer \cite{Gubbins2007,Olson2017,Yan2018,Gastine2020}.  

However, the fluid dynamics of stably-stratified layers has not been fully considered in the stable top layer scenario so far. 
In particular, the assumption that the vertical motions are negligible in the layer might not be valid.
Indeed, the density is affected by at least two components -- the temperature and the composition of the fluid (\eg the concentration of elements lighter than iron such as sulfur or silicon
in the case of terrestrial planets). Importantly, these two components have very different molecular diffusivities, 
which enables a physical process known as double-diffusive convection (DDC) to take place in stable layers \cite{Turner1985,Radko2013}.
DDC can occur if either the temperature gradient or the composition gradient is unstable, while the density gradient remains stable. 
In DDC, the potential energy associated with the destabilising component is released through the rapid molecular diffusion of temperature 
\cite{Stern1960}, and the resulting vertical flows can significantly enhance the transport of heat and composition in stable layers \cite{Traxler2011}. 
DDC has been extensively studied in oceanography \cite{Schmitt1994,Kunze2003}, magma chambers \cite{Sparks1984,Hansen1990} and astrophysics \cite{Garaud2018},
but has only been investigated by a few studies in the context of planetary cores \cite{Manglik2010,Net2012,Bouffard2017,Monville2019,Silva2019,Mather2021}. 
Yet, if present, DDC would distort the magnetic field passing through the layer, thereby affecting the observable magnetic field. 
How DDC flows interact with magnetic fields is largely unknown \cite{Harrington2019}: they could amplify the magnetic field \cite{Manglik2010,Mather2021}
or reduce it through local cancellations or enhanced turbulent diffusion. 
The presence of DDC could therefore invalidate or strengthen the stable top layer scenario. 
The goal of this paper is to describe the dynamics of DDC in conditions relevant to planetary cores.
Most of the aforementioned studies of DDC in planetary cores focus on the instability onset \cite{Net2012,Monville2019,Silva2019}
or consider only a few simulations across the parameter range \cite{Manglik2010,Bouffard2017},
which limits the possibility to extrapolate the results to relevant parameters. 
Following on from the studies of \citeA{Monville2019,Mather2021}, we survey a wide range of the parameter space, but consider 
stronger composition gradients than \citeA{Mather2021} and a spherical shell rather than the full sphere geometry of \citeA{Monville2019}.
Given the richness of the dynamics, as a first step, our model does not include magnetic fields. 
By describing how the properties of DDC (such as typical flow lengthscale and velocity) vary with the parameters (particularly the stratification), 
we attempt to predict its effects on planetary magnetic fields. 

Stable top layers where one component of the density is destabilising are relevant to a variety of planets.
For Mercury, thermal evolution models suggest that the heat flux is subadiabatic in the upper core \cite{Hauck2004}, while a destabilising composition gradient
can form in this region as light elements
are released upwards, due to either the inner core solidification or an underlying iron-snow layer \cite{Dumberry2015}.
This situation is also relevant for Ganymede \cite{Hauck2006}. 
This configuration (stable temperature gradient and unstable composition gradient) is prone to a type of DDC known as fingering convection,
where the primary instability takes the form of vertical plumes or ``fingers" \cite{Stern1960,Turner1985}. 
For Saturn, the immiscibility of helium with hydrogen in the upper part of the metallic layer would result in a downward segregation in the form of a helium rain
and a stable compositional stratification \cite{Stevenson1980}. This configuration (stable composition gradient and unstable temperature gradient) is prone to another 
type of DDC known as oscillatory double-diffusive convection (ODDC) or semi-convection, where the primary instability consists of gravity waves \cite{Baines1969}. 
In the Earth's core, either configuration is plausible depending on thermal evolution scenarios and light element enrichment mechanisms:
a stable temperature gradient could form due to the heat flux becoming locally subadiabatic \cite{Labrosse2015,Greenwood2021} or 
a stable composition gradient could form due to the accumulation of light elements at the top or the core \cite{Buffett2010,Gubbins2013,Landeau2016,Bouffard2019}.
In this paper, we will only consider the case of fingering convection and defer the case of ODDC to a forthcoming study.  

Fingering convection relies on diffusive processes, so the fingers appear on small scales \cite{Stern1960}.
Nevertheless small-scale fingers could affect the magnetic field passing through the stable top layer by enhancing the turbulent magnetic diffusion locally
or inducing magnetic fields.
Additionally, fingering convection can create large density fluctuations, leading to the development of 
secondary instabilities on much larger scales \cite{Stellmach2011,Radko2013}. These are of great interest because they could generate stronger magnetic induction effects. 
One of the best known large-scale structures associated with fingering convection are thermo-compositional staircases:
these are persistent well-mixed layers separated by stratified interfaces, which can be coherent over large distances.
Thermohaline staircases are famously observed in the ocean, in the temperature-salt double-diffusive system
\cite{Schmitt1994}.
However, \citeA{Traxler2011b} showed that the main mechanisms leading to the formation of staircases are ineffective at low Prandtl numbers 
($\Pran$, the ratio of the fluid viscosity to the thermal diffusivity), which is the situation relevant for stellar interiors and 
planetary cores (while oceans have $\Pran \gtrsim 1$). Nevertheless, \citeA{Brown2013} found that staircases form in a limited range of the parameter space,
for very weakly stratified layers (\ie when the density ratio is close to unity, as defined below). 
The relevance of staircases for planetary cores might therefore be limited, but this clearly requires investigation.
In the presence of rotation, other interesting large-scale structures have recently been observed in planar fingering convection in the form of vortices \cite{Sengupta2018}, 
similar to the large-scale vortices observed in turbulent rotating convection \cite{Guervilly2014,Favier2014}.
Most studies of DDC consider local planar domains, where the large-scale flows are confined 
by the computational box size, so the saturation size and amplitude of these flows are unknown and can only be determined 
in a global spherical geometry. 
In this work, we consider the combined effects of spherical geometry and rotation. We are particularly interested in the formation of large-scale flows and global circulation (such as differential rotation). 

DDC is difficult to model numerically because it requires to model diffusivities (viscosity $\nu$, thermal diffusivity $\kappa_t$ and compositional diffusivity $\kappa_c$) 
that have very different values, hence the requirement to compute a wide range of time and length scales. 
In planetary cores, $\kappa_t>\nu>\kappa_c$, with $\kappa_t/\kappa_c =\mathcal{O}(10^3)$ (this ratio is often called the Lewis number $\Le$) 
and $\Pran=\nu/\kappa_t=\mathcal{O}(0.1)$ \cite{Braginsky1995}. 
In numerical simulations of overturning convection, these diffusivities are often set to be the same and the composition and temperature fields are combined together in a codensity variable \cite{Braginsky1995}. The codensity simplification allows to shorten the modelled scale range and to solve only one evolution equation for the codensity variable, 
thereby reducing the computational load.
Numerical studies of rotating spherical convection that use separate evolution equations for the composition and temperature include \citeA{Glatzmaier1996,Breuer2010,Trumper2012,Takahashi2014,Takahashi2019,Tassin2021}, 
but all these studies considered ``top-heavy" configurations (\ie the density gradient is unstable to overturning convection). 
The codensity simplification is clearly not possible in DDC, since the differing diffusivity values is essential to the DDC process.  
As detailed later in this paper, the range of background density gradients unstable to fingering convection increases with $\Le$, so 
this range is expected to be widespread in planetary cores, and we must consider values of $\Le \gg 1$. 
Furthermore, the background density gradients in the stable layer are highly uncertain for any planet, so we need to conduct an extensive parameter
survey to determine its effect on the dynamics.  
Consequently, to accommodate the computational constraints,
our study focuses solely on the dynamics of the stable top layer, neglecting penetrative convection and convective overshoot from the deep convective layer.
In this paper, we consider a thick ``Mercury-like" stable layer, with thickness of 40\% of the core radius $r_o$ (\ie approximately 800km) \cite{Wardinski2021}.

The structure of the paper is as follows. The mathematical formulation of the model is described in Section~\ref{sec:model}. 
The results are presented in Section~\ref{sec:results}, including the variations of the typical length and velocity of fingering convection
with the stratification and rotation rate, the formation of differential rotation, and the efficiency of the convective transport. 
In Section~\ref{sec:discussion}, we discuss the significance of our results   
for planetary magnetic fields. Finally, concluding remarks are contained in Section~\ref{sec:conclusion}.

\section{Model}
\label{sec:model}

\subsection{Governing equations}

\begin{figure}[h]
	\centering
	\includegraphics[clip=true,width=0.5\textwidth]{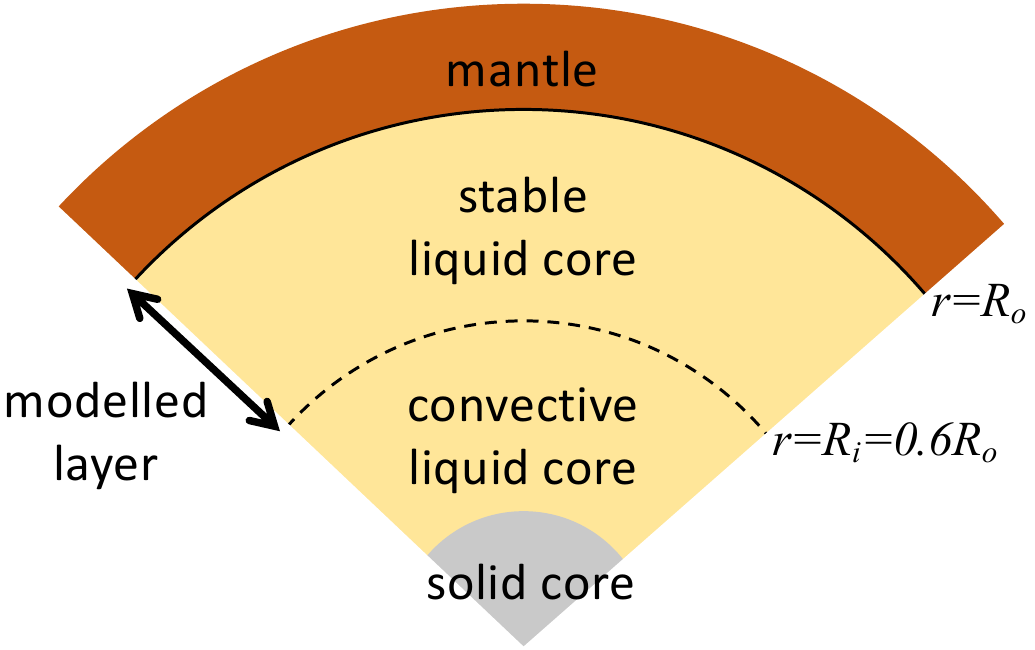}
	\caption{Schematic of the model geometry. The numerical model only includes the stable liquid core layer. 
	The core-mantle boundary at \mbox{$r=r_o$} is assumed to be no-slip and impenetrable for the velocity, with fixed heat flux and zero flux for the composition.
	The bottom boundary representing the interface with the convective liquid core at \mbox{$r=r_i$} is assumed to be stress-free and impenetrable for the velocity,
	with fixed heat and composition fluxes.}
	\label{fig:sketch}
\end{figure}

We model a spherical layer containing a Boussinesq fluid. A schematic of the model is shown in Figure~\ref{fig:sketch}.
The layer is bounded by an inner spherical boundary at radius 
$R_i$, which models the 
interface with the outer convective core, and an outer spherical boundary at radius 
$R_o$, which models the interface with the mantle. 
The layer thickness is fixed to 40\% of the core radius 
$R_o$, and so the aspect ratio is 
$\asp=R_i/R_o=0.6$. 
The layer rotates about the $z$-axis at the rotation rate $\Omega$.
The gravity is directed along the spherical radius ($r$) and is linear in $r$, 
$\vect{g}=-g_o(r/R_o)\vect{e}_r$.

In our Boussinesq system, the equation of state is
\begin{linenomath*}
\begin{equation}
	\frac{\rho}{\rho_m} =  1- \alpha_t (T-T_m) - \alpha_c(C-C_m),
\label{eq:state}
\end{equation}
\end{linenomath*}
where $\rho$ is the density, $T$ the temperature and $C$ the concentration in light elements (which we will call the ``composition"), 
$\alpha_t$ the coefficient of thermal expansion, and $\alpha_c$ the compositional analogue to $\alpha_t$.
The subscript $m$ denotes a constant mean value. 
The coefficients $\alpha_t$ and $\alpha_c$ are positive, \ie positive deviations of temperature or composition from the mean value 
create relative density deficits.
The temperature and composition are decomposed into a constant mean, a static (or background) profile (denoted by the subscript $s$) and a perturbation:
\begin{linenomath*}
\begin{equation}
	T = T_m+T_s + \Tpert, \quad C = C_m+C_s +\Cpert.
\end{equation}
\end{linenomath*}

The fluid has viscosity $\nu$, thermal diffusivity $\kappa_t$ and compositional diffusivity $\kappa_c$, all of which are constant, with $\kappa_t>\nu>\kappa_c$. 
The diffusivity coefficients are fixed throughout this study. Following \citeA{Monville2019}, we set the Prandtl number
$\Pran=\nu/\kappa_t=0.3$ and the Schmidt number $\Sc=\nu/\kappa_c=3$. 
This implies that the Lewis number, defined as $\Le=\kappa_t/\kappa_c=\Sc/\Pran$, is equal to $10$ in all the simulations.
Though smaller than planetary core values ($\Le=\mathcal{O}(10^3)$), 
this choice of $\Le$ is computationally achievable and still provides a substantial
parameter space to study double-diffusive fingering instabilities (see below Equation~\eqref{eq:range}).

The system of governing equations is solved in a dimensionless form, where the unit for length is the layer thickness 
$D=R_o-R_i$ 
(\ie in dimensionless form, $r_i=R_i/D=1.5$ and $r_o=R_o/D=2.5$).
The unit for time is $D^2/\nu$, for temperature $\nu^2/\alpha_t g_o D^3$, and for composition  $\nu^2/\alpha_c g_o D^3$. 
The Navier-Stokes equation in dimensionless form is thus
\begin{linenomath*}
\begin{equation}
	\pd{\vel}{t} + (\vel \cdot \vn )\vel + \frac{2}{\Ek}\vect{e}_z \times \vel = -\vn p + \vn^2 \vel + \pl \Tpert + \Cpert \pr  \frac{r}{r_o} \vect{e}_r,
	\label{eq:NS}
\end{equation}
\end{linenomath*}
where $\vel$ is the solenoidal velocity (\ie $\vn \cdot \vel =0$), $p$ the pressure, and all the variables are now dimensionless. 
The Ekman number is a dimensionless number defined as 
\begin{linenomath*}
\begin{equation}
	\Ek = \frac{\nu}{\Omega D^2}.
\end{equation}
\end{linenomath*}

The governing equations for the temperature and composition perturbations are
\begin{linenomath*}
\begin{eqnarray}
	\pd{\Tpert}{t} + \vel \cdot \vn \Tpert  + u_r \frac{\textrm{d} T_s}{\textrm{d}r}  &=&  \frac{1}{\Pran} \n^2 \Tpert, \label{eq:Tpert}
	\\
	\pd{\Cpert}{t} + \vel \cdot \vn \Cpert + u_r \frac{\textrm{d} C_s}{\textrm{d}r}  &=&  \frac{1}{\Sc} \n^2 \Cpert, \label{eq:Cpert}
\end{eqnarray}
\end{linenomath*}
where $T_s=T_s(r)$ and $C_s=C_s(r)$.
The radial gradients of the background fields are obtained by solving the diffusion equations 
\begin{linenomath*}
\begin{eqnarray}
	 \frac{1}{\Pran} \n^2 T_s &=& 0,
	 \\
	 \frac{1}{\Sc} \n^2 C_s &=& -\epsilon,
	 \label{eq:Cs}
\end{eqnarray}
\end{linenomath*}
where $\epsilon$ is a constant.
We fix the flux of temperature and composition at the bottom boundary,
\begin{linenomath*}
\begin{eqnarray}
	\left.  \frac{\textrm{d} T_s}{\textrm{d}r} \right|_{r_i} = - q_i  \frac{\alpha_t g_o D^4}{\nu^2}, \quad \left.  \frac{\textrm{d} C_s}{\textrm{d}r} \right|_{r_i} = - f_i \frac{\alpha_c g_o D^4}{\nu^2}.
\end{eqnarray}
\end{linenomath*}
For fingering convection, where the temperature gradient is stable and the compositional gradient is unstable, we have $q_i<0$ and $f_i>0$.
There is no source term in the equation for the static background temperature, 
so the outward heat flux integrated over the outer boundary equals the incoming heat flux integrated over the inner boundary to ensure a steady state.
We assume that there is no flux of light elements to/from the mantle at $r=r_o$, hence $\textrm{d}C_s/\textrm{d}r=0$ at $r=r_o$. 
The term $\epsilon$ in equation~\eqref{eq:Cs} is a sink term that compensates the inflow of light elements at the lower boundary to achieve a steady state
and is related to $f_i$ by
\begin{linenomath*}
\begin{eqnarray}
	\epsilon = - \frac{3}{\Sc} f_i \frac{\alpha_c g_o D^4}{\nu^2}  \frac{r_i^2}{r_o^3-r_i^3}.
\end{eqnarray}
\end{linenomath*}
The gradients of the background fields in the domain are thus 
\begin{linenomath*}
\begin{eqnarray}
	\quad \frac{\textrm{d} T_s}{\textrm{d}r} &=&  - \frac{\Ra_t}{\Pran} \frac{r_i}{r^2},
	\\
	\frac{\textrm{d} C_s}{\textrm{d} r} &=& \frac{Ra_c}{\Sc(1-\asp^3)}  \pl \frac{r_i }{r_o^3} r - \frac{r_i}{r^2} \pr,
\end{eqnarray}
\end{linenomath*}
and the Rayleigh numbers are dimensionless numbers defined as
\begin{linenomath*}
\begin{equation}
	\Ra_t = \frac{\alpha_t g_o q_i R_i D^3}{\nu \kappa_t}, \quad \Ra_c = \frac{\alpha_c g_o f_i R_i D^3}{\nu \kappa_c}.
\end{equation}
\end{linenomath*}
For fingering convection, we have $\Ra_t<0$ and $\Ra_c>0$.

The boundary conditions for the perturbations are zero radial fluxes at $r=r_i$ and $r=r_o$,
\begin{linenomath*}
\begin{eqnarray}
	\left. \pd{\Tpert}{r} \right|_{r_i} = \left. \pd{\Cpert}{r} \right|_{r_i} = \left. \pd{\Tpert}{r} \right|_{r_o} = \left. \pd{\Cpert}{r} \right|_{r_o} = 0.
\label{eq:BC_TC}
\end{eqnarray}
\end{linenomath*}

For the velocity, we use impenetrable boundary conditions at $r=r_i$ and $r=r_o$, stress-free boundary conditions at $r=r_i$ to model the interface with the convective liquid core, 
and no-slip boundary conditions at $r=r_o$ to model the core-mantle boundary.
The choice of impenetrable boundary condition at $r=r_i$ is made for numerical convenience since the underlying convective layer is not included in the model. 
Any dynamical interaction between the convective and stable layers (such as convective overshoot and penetration) is thus ignored in our model for simplicity.

\subsection{Buoyancy frequency of the background state}

\begin{figure}
	\centering
	\includegraphics[clip=true,width=\textwidth]{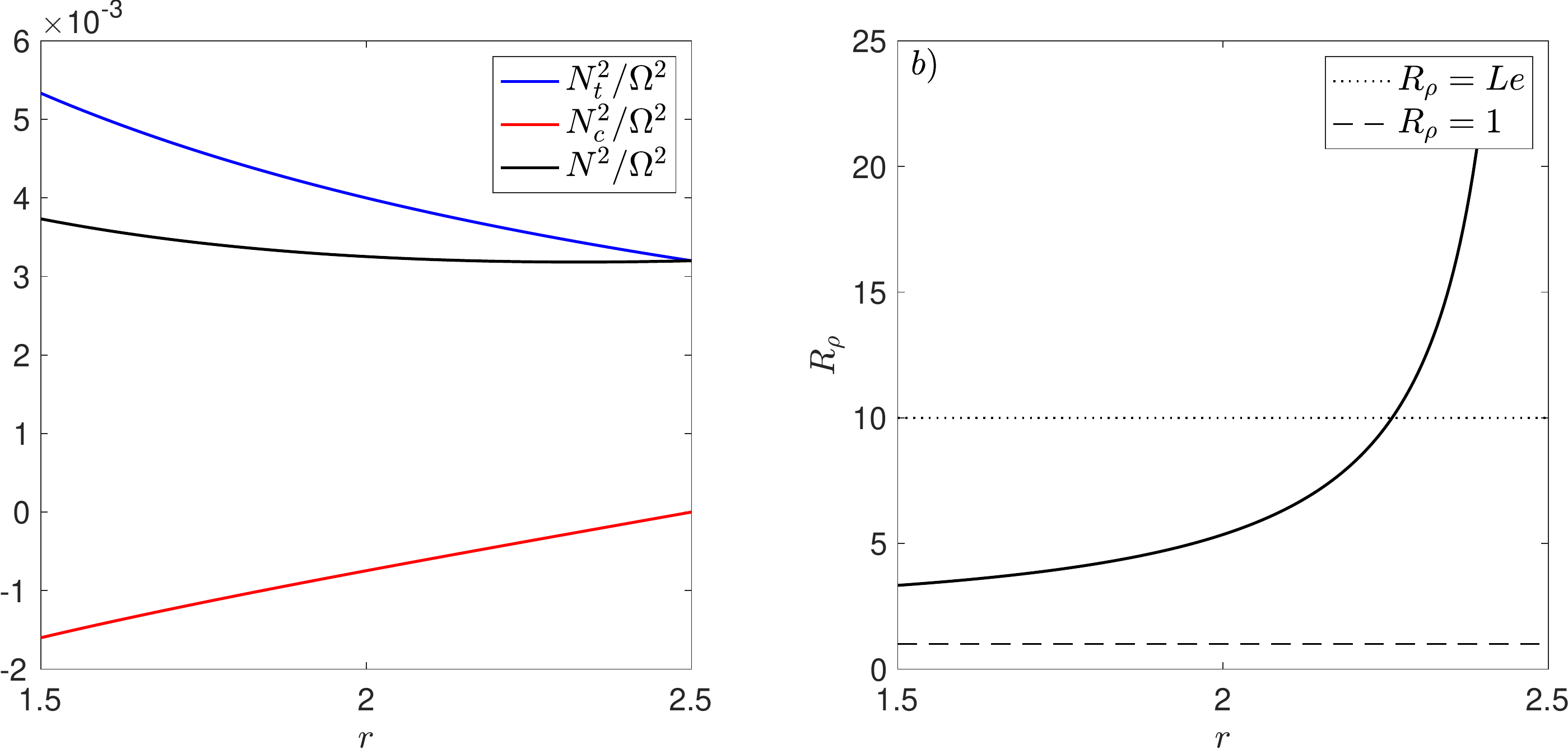}
	\caption{Radial profiles of (a) the buoyancy frequencies associated with the background temperature gradient, $N_t^2$, the background composition gradient, $N_c^2$,
	and the background density gradient, $N^2$, normalised by $\Omega^2$, and 
	(b) the density ratio $R_{\rho}$ for \mbox{$\Ek=10^{-5}$}, 
	\mbox{$\Ra_c=1.2\times 10^8$}, 
	\mbox{$\Ra_t=-\Ra_c/3$}, \mbox{$\Pran=0.3$} and \mbox{$\Sc=3$}.}
	\label{fig:N2}
\end{figure}

The buoyancy frequency (or Brunt-V\"ais\"al\"a frequency) corresponds to the frequency 
of oscillations of fluid parcels displaced in the vertical direction in stably stratified systems and is used to measure the stability of a fluid to vertical motions.
In the Boussinesq approximation, the buoyancy frequency is
\begin{linenomath*}
\begin{equation}
	N^2 = \vect{g}\cdot \n \pl \frac{\rho}{\rho_m} \pr.
\end{equation} 
\end{linenomath*}
In a stably-stratified layer, the radial gradient of density is negative, so $N^2>0$.  
The system is unstably stratified when $N^2<0$ and neutral when $N^2=0$. $N^2$ is often used as a measure of the strength of the stratification. 
Using the equation of state~\eqref{eq:state}, $N^2$ can be decomposed into a thermal component $N_t^2$ associated with the radial gradient
of $T_s$ and a compositional component $N_c^2$ associated with the radial gradient of $C_s$, 
\begin{linenomath*}
\begin{equation}
	N^2 = N_t^2 +N_c^2. 
\end{equation} 
\end{linenomath*}
In the fingering convection system, $N^2_c<0$, and $N^2_t>0$.
With our choice of units, in dimensionless form, these become 
\begin{linenomath*}
\begin{equation}
	\tilde{N}^2_t = \frac{N_t^2}{\nu^2/D^4} = \frac{r}{r_o} \frac{\textrm{d}T_s}{\textrm{d}r} \quad  \textrm{and} \quad  
	\tilde{N}^2_c = \frac{N_t^2}{\nu^2/D^4} =  \frac{r}{r_o} \frac{\textrm{d}C_s}{\textrm{d}r}.
\end{equation} 
\end{linenomath*}

At the lower and upper boundaries, we have 
\begin{linenomath*}
\begin{eqnarray}
	&& \tilde{N}^2_t(r_i) = - \frac{\Ra_t}{\Pran}\frac{1}{r_o}, \quad  \tilde{N}^2_c(r_i) = - \frac{\Ra_c}{\Sc}\frac{1}{r_o},
	\\
	&& \tilde{N}^2_t(r_o) =  \tilde{N}^2_t(r_i) \asp, \quad  \tilde{N}^2_c(r_o) = 0.
\end{eqnarray}
\end{linenomath*}

$N^2$ is commonly normalised by $\Omega^2$ to compare the effects of the stratification with the rotational effects, in which case
\begin{linenomath*}
\begin{equation}
	\frac{N^2}{\Omega^2} = \Ek^2 \pl \tilde{N}^2_t + \tilde{N}^2_c \pr.
\end{equation}
\end{linenomath*}

Figure~\ref{fig:N2}a shows the radial profiles of $N^2$, $N_t^2$ and $N_c^2$ normalised by $\Omega^2$ for 
$\Ek=10^{-5}$, 
$\Ra_c=1.2\times 10^8$, and 
$\Ra_t=-4\times10^7$. 
The total stratification is strongest at the bottom of the layer, but the variations of $N^2$ across the layer are relatively modest with 
$N^2(r_i)/N^2(r_o)=(1+\Ra_c/(\Le \Ra_t))/\asp=1.17$ for $\Ra_t=-\Ra_c/3$, which corresponds to a set of parameters that we will analyse in detail in this paper.

\subsection{Regime unstable to fingering convection}
\label{sec:onset}

The domain of stability for double-diffusive instability is defined using the density ratio $R_{\rho}=|N_t^2|/|N_c^2|$.
The system is stable to overturning convection when $N^2>0$, i.e. $N_t^2>-N_c^2$ or $R_{\rho}>1$.
In non-rotating planar layers with constant $N_t^2$ and $N_c^2$, the system is prone to the double-diffusive fingering instabilities when \cite{Stern1960}
\begin{linenomath*}
\begin{equation}
	1<R_{\rho}<\Le.
	\label{eq:range}
\end{equation}
\end{linenomath*}
This unstable range is also valid for rotating planar systems and the fastest growing modes (the so-called elevator modes,
which span the whole layer depth for unbounded gradient layers) are unaffected by rotation, while the growth rate of other modes is reduced by the rotation \cite{Sengupta2018}.
Studies of the onset of rotating fingering convection in spherical geometry find that
large-scale modes (with low azimuthal wavenumber) 
are preferred near the edges of the stability domain and form ``pockets" of instability for small $\Ra_c$  \cite{Silva2019,Monville2019}. 
For large $\Ra_c$, the stability curves collapse onto the regime of non-rotating fingering convection~\eqref{eq:range} \cite{Monville2019}.

In our study, the density profile $R_{\rho}$ varies with radius. The radial profile of $R_{\rho}$ is shown on Figure~\ref{fig:N2}b for \mbox{$\Ra_t=-\Ra_c/3$} and $\Le=10$. 
At $r_i$, \mbox{$R_{\rho}(r_i)=\Le |\Ra_t|/\Ra_c$}, and so, at the bottom of the domain, the unstable range~\eqref{eq:range}  corresponds to
\begin{linenomath*}
\begin{equation}
	\Ra_c/\Le < |\Ra_t| < \Ra_c.
	\label{eq:range2}
\end{equation}
\end{linenomath*}
$R_{\rho}$ increases with radius, going to infinity at $r=r_o$ as $N_c^2(r_o)=0$. This implies that, while the bottom of the domain
might be prone to fingering instabilities, the top of the domain is not. The line $R_{\rho}=\Le$ is crossed in the domain at a ``stable" radius $r_s$. 
As the radial profile of $R_{\rho}$ depends on $|\Ra_t|/\Ra_c$ (and not on $\Ra_t$ or $\Ra_c$ individually), 
fixing this ratio leaves the stable radius unchanged.
For \mbox{$\Ra_t=-\Ra_c/3$}, $r_s=2.26$ so $76\%$ of the radial domain is in the region where we would expect fingering convection to occur
at large Rayleigh numbers.

\subsection{Numerical method}
All the simulations are performed with the open-source code XSHELLS (\url{https://nschaeff.bitbucket.io/xshells/}) \cite{Schaeffer2013}.
XSHELLS is a C++ pseudo-spectral code that solves the governing equations \eqref{eq:NS}, \eqref{eq:Tpert} and \eqref{eq:Cpert} in a 3D spherical geometry. 
The velocity is decomposed into poloidal and toroidal scalars,
which ensures that the velocity is solenoidal. All the scalars (poloidal and toroidal velocity, temperature and composition perturbations)
are expanded in spherical harmonics $Y_l^m(\theta,\phi)$ of degree $l$ and order $m$, which are truncated at $L_{max}$ and $M_{max}$ respectively.
An azimuthal symmetry (either 2-fold or 4-fold) is imposed for the most computationally demanding simulations (the two largest 
Rayleigh numbers at $\Ek=10^{-5}$ and most of the simulations at $\Ek=10^{-6}$).
In the radial direction, the code uses a second-order finite difference scheme with $N_r$ points. 
Details of the numerical resolution and the simulated time (in units of the viscous timescale)
used for each simulation are provided in the dataset available on the Newcastle University Research Data Repository \cite{dataset}.
The code uses a second-order time-stepping scheme with an implicit treatment of the diffusive terms and explicit treatment of the non-linear terms.  
XSHELLS was benchmarked against codes used in the geodynamo community \cite{Marti2014,Matsui16} and has been previously used to study DDC 
in a full sphere geometry \cite{Monville2019}.

\section{Results}
\label{sec:results}

\subsection{Survey of the parameter space}
\label{sec:survey}

Most of the simulations were performed at $\Ek=10^{-5}$, with some additional simulations at $\Ek=10^{-4}$ and $\Ek=10^{-6}$
to assess the effect of varying the rotation rate. 
To reduce the degrees of freedom in the parameter space, our study mainly focusses on a fixed density ratio corresponding to
\mbox{$\Ra_t=-\Ra_c/3$} (\ie \mbox{$R_{\rho}(r_i)=\Le/3$}).
Additionally, we have performed a number of simulations at fixed $\Ra_t$ 
($\Ra_t=-4\times10^7$) to study the dependence of the flow 
on the density ratio as $R_{\rho}(r_i)$ varies from $1$ to $\Le$.

\begin{figure}[h!]
	\centering
	\includegraphics[clip=true,width=\textwidth]{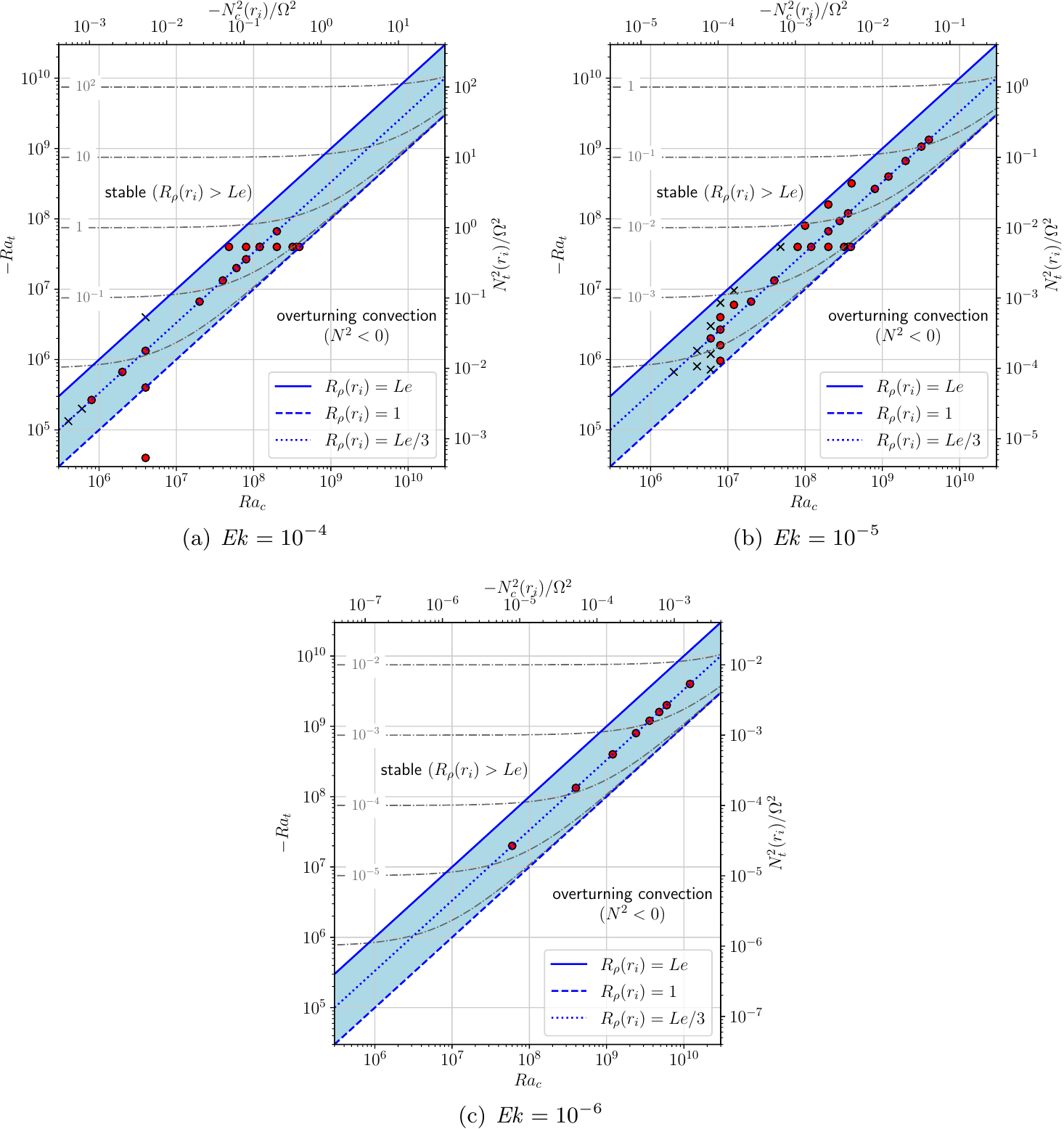}
	\caption{Simulations in the parameter space \mbox{$|\Ra_t|-\Ra_c$} (or equivalently \mbox{$N_t^2(r_i)/\Omega^2$ }and 
	\mbox{$|N_c^2(r_i)|/\Omega^2$} as indicated on the right and top axis) for three Ekman numbers. 
	The crosses are the cases stable to fingering convection (\ie where the kinetic energy decays). 
	The blue area represents the parameter range \mbox{$1<R_{\rho}(r_i)<\Le$}.
	The gray dash-dotted lines represent isolines of \mbox{$N^2(r_i)/\Omega^2$}.}
	\label{fig:Le10}
\end{figure}

Figure~\ref{fig:Le10} shows the location of our simulations in the parameter space \mbox{$|\Ra_t|-\Ra_c$} for $\Ek=\{10^{-4},10^{-5},10^{-6}\}$.
The crosses represent cases where the kinetic energy decays; the circles represent cases where the kinetic energy initially grows (after 
an initial condition of small amplitude is applied) and subsequently saturates.  
The blue area corresponds to the parameter range for which the bottom of the domain might be prone to the fingering instability
according to Equation~\eqref{eq:range2}. 
We find that the minimum value of $\Ra_c$ at which fingering convection onsets is 
$8\times10^5$ for $\Ek=10^{-4}$ and 
$6\times10^6$ for $\Ek=10^{-5}$. 
These values can be compared with the onset of overturning compositional convection (\ie at $\Ra_t=0$), 
which we find to be located at 
$\Ra_c\approx8.8\times10^5$ for $\Ek=10^{-4}$ and 
$\Ra_c\approx2.1 \times 10^7$ for $\Ek=10^{-5}$. 
Consequently, fingering convection onsets earlier (\ie at smaller $\Ra_c$) than overturning compositional convection, especially at small $\Ek$,
in agreement with the results of \citeA{Monville2019,Mather2021}.
\citeA{Monville2019} found that the minimum $\Ra_c$ required for the onset of fingering convection scales as $\Ek^{-1}$ for small $\Ek$ and depends on $\Le$ (and not on $\Pran$
and $\Sc$ individually), meaning that the onset of rotating fingering convection at low Rayleigh numbers is independent of viscosity. 
Here we find a similar scaling for the minimum $\Ra_c$, although a slightly smaller exponent (approximately $-0.9$), is a better fit to our data, meaning that
the simulations performed at $\Ek=\{10^{-5},10^{-4}\}$ are still not quite in the inviscid regime described by \citeA{Monville2019}.

\subsection{3D structure of fingering convection}
\label{sec:3D}

\begin{figure}[h]
	\centering
	\includegraphics[clip=true,width=\textwidth]{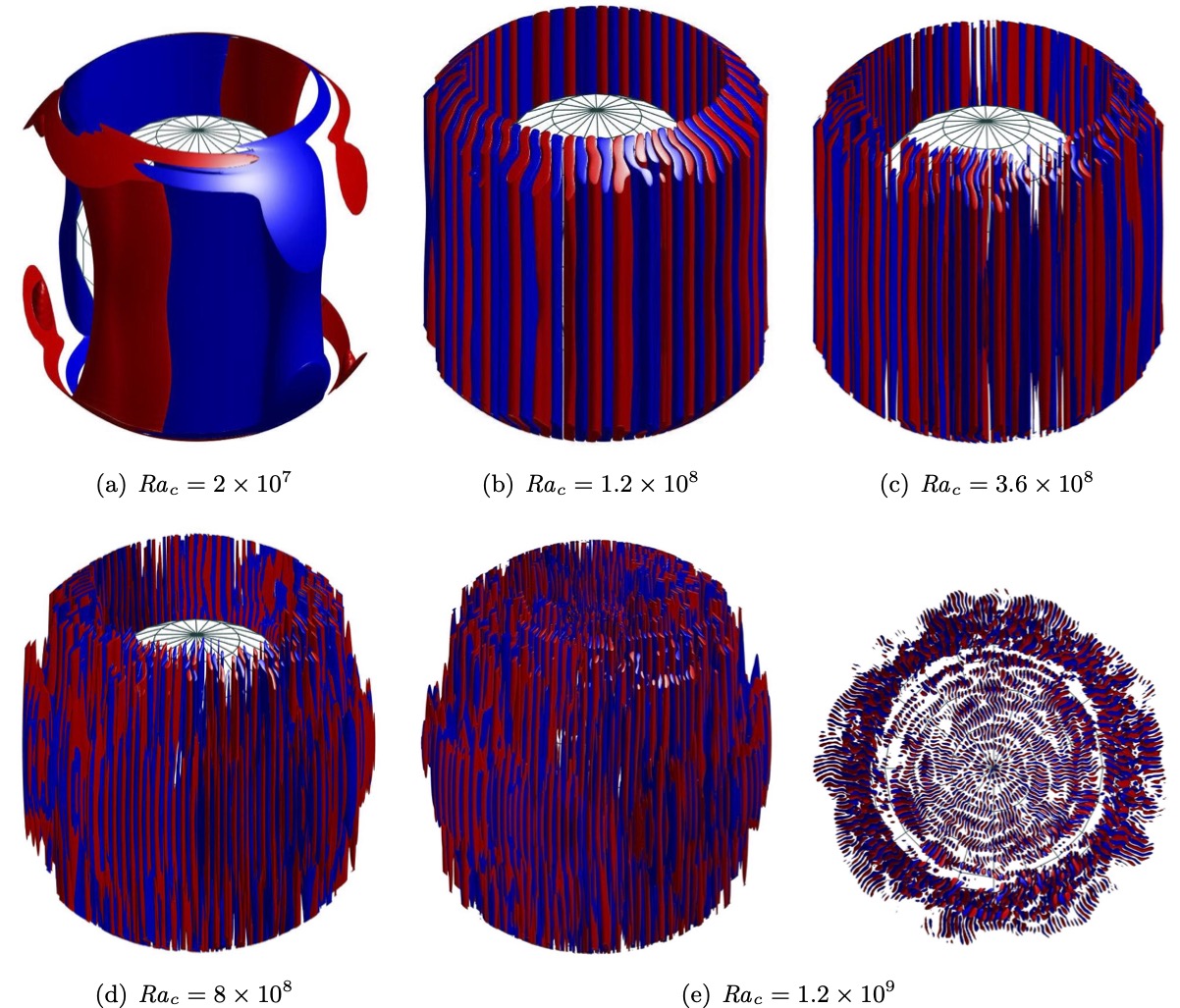}
	\caption{Isosurfaces of $u_s$ for \mbox{$\Ek=10^{-5}$} and \mbox{$\Ra_t=-\Ra_c/3$}. 
	For 
	\mbox{$\Ra_c=2\times10^7$} (a), the snapshot is taken at \mbox{$t=140$} (see time series of the kinetic energy in Figure~\ref{fig:t_KE_E5Ra5e7}).
	For 
	\mbox{$\Ra_c=1.2\times 10^9$} (e), a polar view is also displayed to 
	highlight the flow inside the tangent cylinder. The isosurfaces corresponds to \mbox{$\pm 15\%$} of the maximum value in each case (red: positive, blue: negative).}
	\label{fig:isosurf}
\end{figure}

\begin{figure}[h]
	\centering
	\includegraphics[clip=true,width=0.7\textwidth]{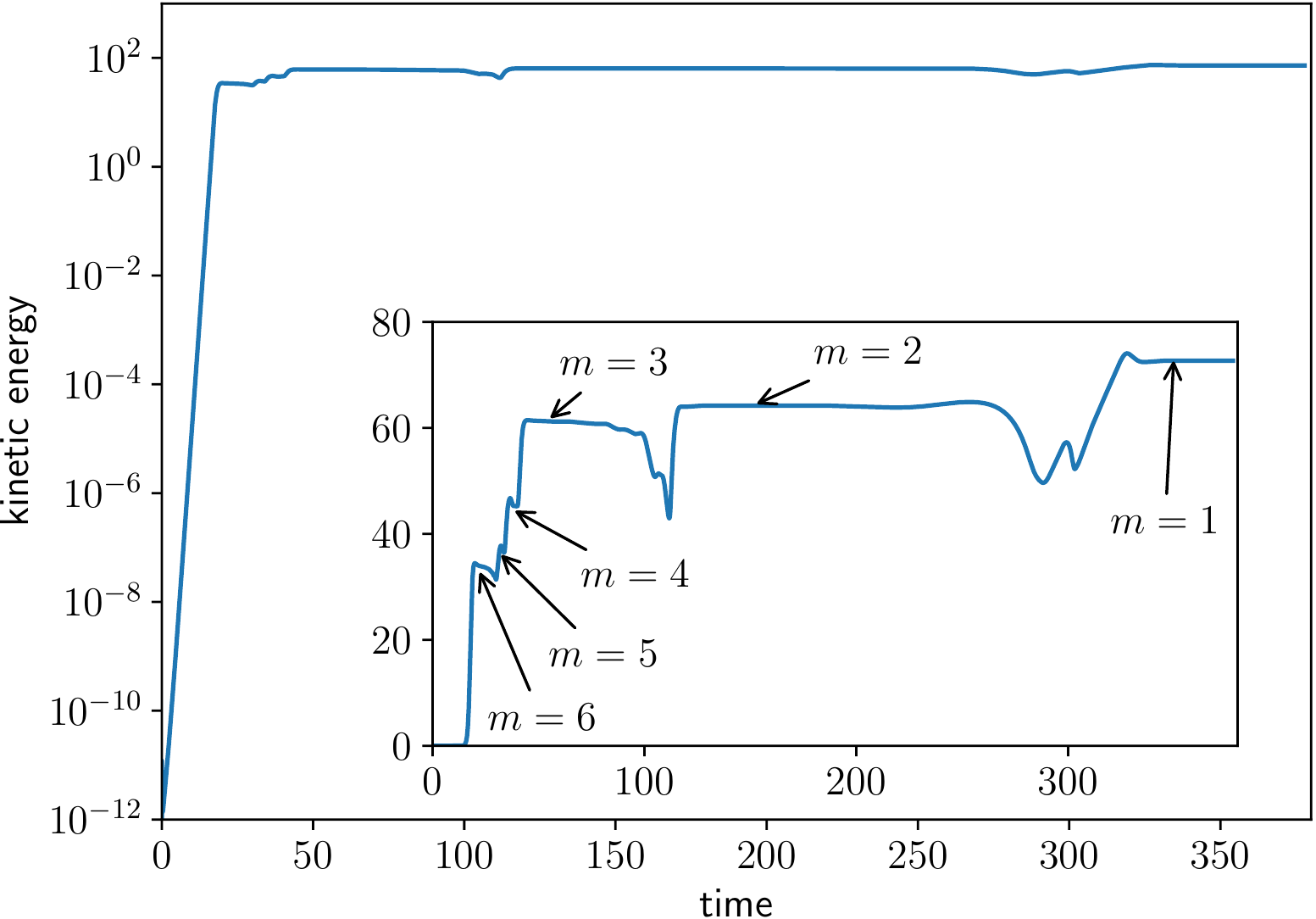}
	\caption{Times series of the kinetic energy for \mbox{$\Ek=10^{-5}$}, \mbox{$\Ra_t=-\Ra_c/3$}, and 
	\mbox{$\Ra_c=2\times10^7$}. 
	The inset shows the same times series with a linear
	scale on the $y$-axis. The annotations on the inset give the preferred azimuthal wavenumber for each phase.}
	\label{fig:t_KE_E5Ra5e7}
\end{figure}

Figure~\ref{fig:isosurf} shows isosurfaces of the cylindrical radial velocity $u_s$ for $\Ek=10^{-5}$, \mbox{$\Ra_t=-\Ra_c/3$} and varying $\Ra_c$.
Near the onset, for 
$\Ra_c=2\times 10^7$, 
the fingering convection takes the form of elongated columns aligned with the rotation axis with an azimuthal length 
comparable with the layer width. The columns are located outside the tangent cylinder (TC).
This snapshot was taken at the time $t=140$, where the dominant azimuthal wavenumber is $m=2$. The dominant wavenumber varies during the saturated phase. 
To illustrate these changes, Figure~\ref{fig:t_KE_E5Ra5e7} shows the time series of the kinetic energy for this simulation. 
This case is located fairly close to the onset, so the growth rate of the instability is slow and 
the kinetic energy saturates after approximately 10 viscous timescales.
After saturation,
the dominant mode is initially $m=6$, but decreases
gradually during the step wise increases of the kinetic energy, until $m=1$ becomes the preferred mode.  

For 
\mbox{$\Ra_c\geq 1.2\times10^8$}, the azimuthal size of the columns becomes much smaller and visibly decreases 
with increasing $\Ra_c$. 
These structures are elongated in the $s$-direction and fairly $z$-invariant, so they appear more sheet-like than finger-like. 
However we will call these structures ``fingers" as they correspond to the fingers of non-rotating planar fingering convection. 
The invariance of the flow along the rotation axis is imposed by the Proudman-Taylor constraint due to the rapid background rotation \cite{Taylor1922}. 
In the equatorial plane, the fingers appear near the lower boundary, where $R_{\rho}$ is smallest (see Figure~\ref{fig:N2}b), 
but, as they extend along the $z$-axis, they cross regions of larger $R_{\rho}$ (\ie weaker background composition gradients)
all the way to the outer boundary, where $R_{\rho}\to\infty$.
 
The preference for large-scale modes near the onset of fingering convection at small Rayleigh numbers and, subsequently, 
the preference for fingers of smaller azimuthal size for larger Rayleigh numbers is consistent with
results obtained in non-rotating planar geometry. In this geometry, cells having horizontal dimension comparable to the layer depth are preferred at the critical onset, 
but smaller fingers have the largest growth rate for large Rayleigh numbers \cite{Stern1960}. 
This observation is also in agreement with the results of \citeA{Monville2019} in spherical geometry. 
As discussed by \citeA{Stern1960}, the radial velocity of wide fingers must be small because the diffusion of their temperature perturbation is slow.  
Thin fingers are preferred at larger Rayleigh numbers because the faster thermal diffusion across the fingers allows greater radial velocities and 
a more efficient release of the potential energy in the background composition gradient.

\begin{figure}[h]
	\centering
	\includegraphics[clip=true,width=\textwidth]{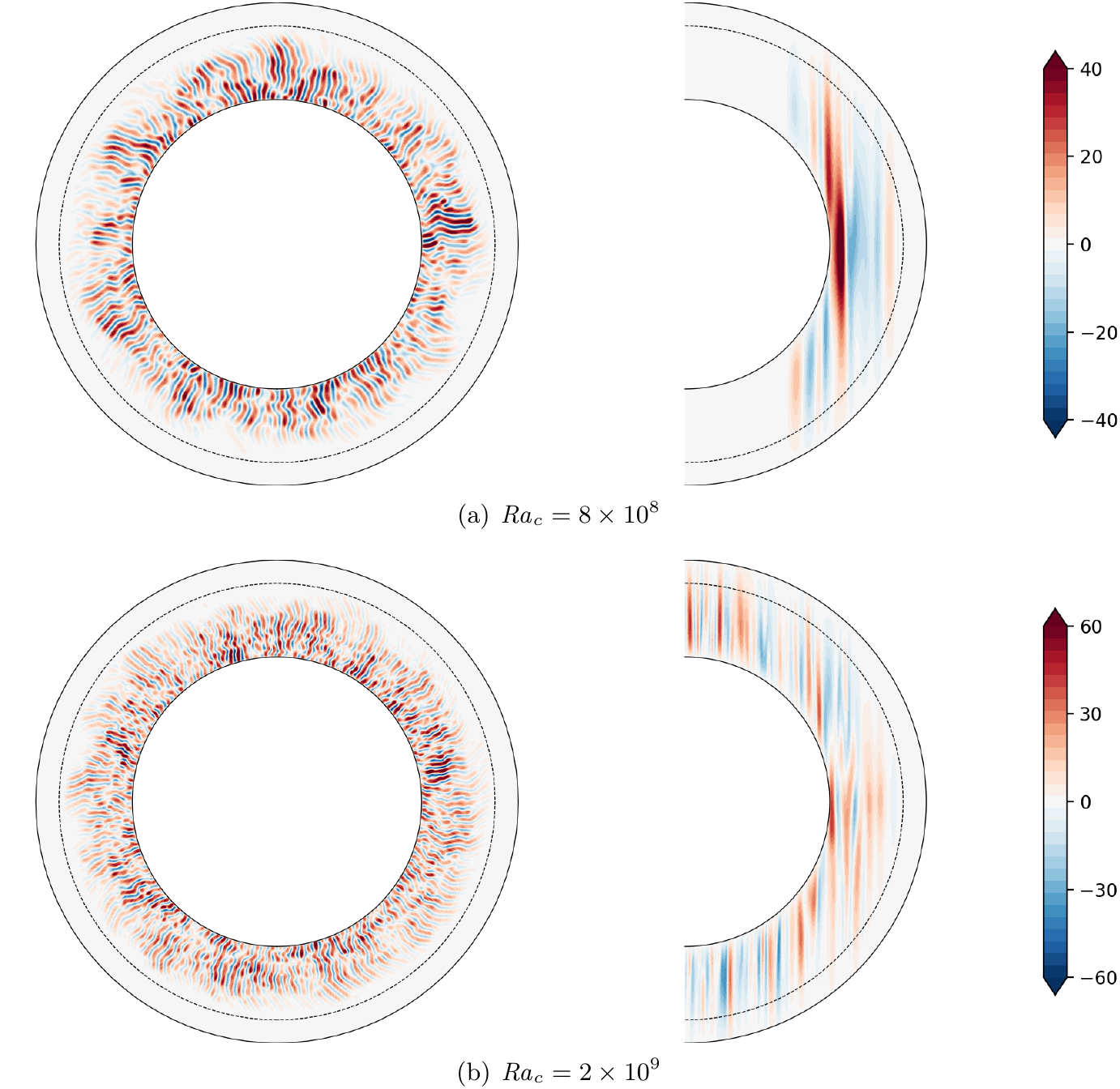}
	\caption{Equatorial and meridional cross-sections of the radial velocity $u_r$ for \mbox{$\Ek=10^{-5}$} and \mbox{$\Ra_t=-\Ra_c/3$}. 
	The dashed line represents \mbox{$r=r_s$}, the radius at which \mbox{$R_{\rho}=\Le$}.}
	\label{fig:equat_merid}
\end{figure}

Figure~\ref{fig:isosurf} indicates that fingers appear inside the TC with a similar structure as the fingers outside the TC at 
$\Ra_c = 1.2\times10^9$. 
The delayed onset of buoyancy instabilities inside the TC is also a well-known feature in standard overturning convection \cite{JonesToG}. 
The inhibiting effect of rotation on the buoyancy instability is stronger when 
the direction of gravity and the rotation axis are aligned \cite{Cha61}, so stronger compositional Rayleigh numbers are required for convection to onset inside the TC.
The beginning of fingering convection inside the TC can be observed in Figure~\ref{fig:equat_merid}, which shows equatorial and meridional cross-sections of the radial velocity $u_r$
for two different $\Ra_c$. 
In both cases, the fingers extend to approximately the stable radius $r_s$ (at which $R_{\rho}=\Le$) in the equatorial plane, 
beyond which the weak background composition gradient cannot maintain fingering convection. 
However, at higher latitudes, the columnar flows poke through the region $r>r_s$, with an amplitude decreasing with increasing height.  
For 
$\Ra_c=8\times10^8$, fingering occurs mostly outside the tangent cylinder, but there is some activity on the inner side of the TC. 
At larger $\Ra_c$, fingering convection occurs at all latitudes inside the TC. 
While the $z$-invariance of the flow degrades outside the TC when $\Ra_c$ increases, the velocity remains $z$-invariant to a good degree inside the TC. 
The only region with no significant radial velocity is the outermost equatorial region where $r>r_s$. 
In addition to large $R_{\rho}$, the slope of the outer boundary is large there, so the vortex stretching produced by columnar flows moving inwards or outwards inhibits the flows
\cite{Guervilly2019}. 

\begin{figure}[h]
	\centering
	\includegraphics[clip=true,width=0.9\textwidth]{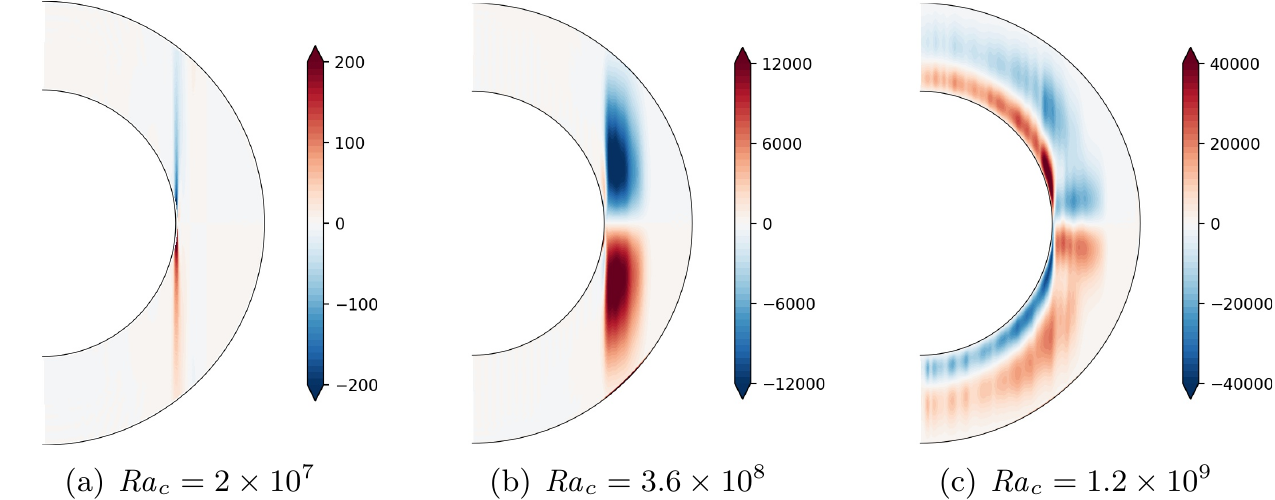}
	\caption{Meridional cross-sections of the axisymmetric kinetic helicity (snapshots)  for \mbox{$\Ek=10^{-5}$} and \mbox{$\Ra_t=-\Ra_c/3$}.}
	\label{fig:helicity}
\end{figure}

To further characterise the 3D structure of the flow, we look at the kinetic helicity, which describes the spatial correlation between the components of the velocity and vorticity 
($\boldsymbol{\omega}=\vn \times \vel$),
\begin{linenomath*}
\begin{equation}
	H = \vel \cdot \boldsymbol{\omega}.
\end{equation}
\end{linenomath*}
The kinetic helicity is often thought to be an essential ingredient in the generation of large-scale magnetic fields \cite{Sreenivasan2011,Moffatt2019}, 
although this idea is contested \cite{Cattaneo2006}. Here we can compare the distribution of the helicity produced by fingering convection with the one
produced by standard overturning convection, whose dynamo properties have been extensively studied \cite{Olson1999,Soderlund2012,ChristensenToG}. 
Figure~\ref{fig:helicity} shows meridional cross-sections of the axisymmetric (\ie azimuthally-averaged) kinetic helicity  for three different $\Ra_c$.
In all cases, the helicity is equatorially antisymmetric (mainly negative in the northern hemisphere) as is the case in rotating spherical overturning convection \cite{Olson1999}.
This antisymmetry follows from the opposite symmetry of the velocity and vorticity (\eg $u_r$ is equatorially symmetric but $\omega_r$ is 
equatorially anti-symmetric).
For 
$\Ra_c=2\times10^7$, where fingering convection consists of large-scale modes, the helicity is confined to a very thin layer on the tangent cylinder. 
At larger $\Ra_c$, where fingering convection takes the form of thin sheet-like structures, the helicity is more broadly distributed within the layer
and follows the location of fingering convection: for 
$\Ra_c=3.6\times10^8$, the helicity is only present outside the TC, 
while it occupies all latitudes for 
$\Ra_c=1.2\times10^9$.
In this case, a layer of positive (negative) helicity appears in the lower northern (southern) region inside the TC as the 
flow changes the direction in which it is spiralling as it moves in/outwards (this also happens in rotating overturning convection \cite{Cha61}).
Since the background temperature and composition gradients depend on radius, the sign change does not occur at the mid-radius of the layer,
as might be expected in Boussinesq system with constant background gradients. 
Overall the distribution of helicity in fingering convection at large Rayleigh numbers is fairly similar to the one produced by overturning convection.

\subsection{Azimuthal finger length}

\begin{figure}[h]
	\centering
	\includegraphics[clip=true,width=\textwidth]{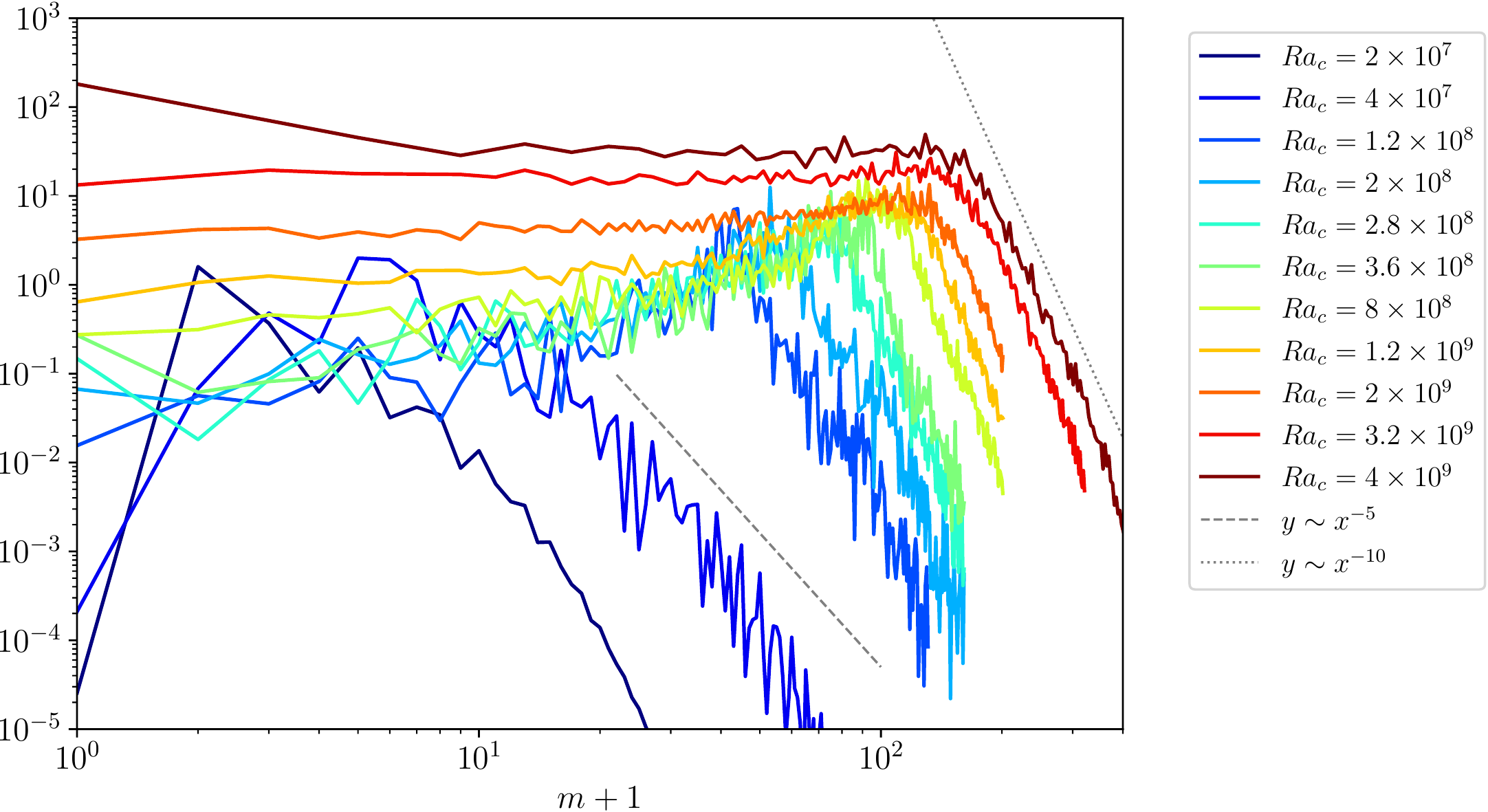}
	\caption{Spectra of the poloidal kinetic energy (snapshot) as a function of the azimuthal order $m$  
	for \mbox{$\Ek=10^{-5}$} and \mbox{$\Ra_t=-\Ra_c/3$}. The spectra are averaged in radius over the whole domain.}
	\label{fig:spectra_E5}
\end{figure}

As noted on Figure~\ref{fig:isosurf}, the azimuthal length of the radial flow visibly decreases with increasing $\Ra_c$ 
(and $\Ra_t$ since we considered the case \mbox{$\Ra_t=-\Ra_c/3$}).
This decrease of the azimuthal finger length is expected from linear theory.
In fingering convection in a non-rotating planar domain, the fastest growing mode (\ie the mode for which the linear instability has the largest growth rate)
at large Rayleigh numbers is expected to have approximately the horizontal length \cite{Stern1960}
\begin{linenomath*}
\begin{equation}
	d = \pl \frac{\kappa_t \nu}{N_t^2}\pr^{1/4}.
	\label{eq:ltheo}
\end{equation}
\end{linenomath*}
This optimal horizontal length allows for thermal diffusion to act effectively 
before the viscosity can suppress the instability. In dimensionless form, $d/D=(\Pran \tilde{N}^2_t)^{-1/4}$, and so, at $r=r_i$,
\begin{linenomath*}
\begin{equation}
	\frac{d}{D} = \pl \frac{r_o}{|\Ra_t |}\pr^{1/4}.
\end{equation}
\end{linenomath*}
In practice, the horizontal length of the fastest growing mode also depends on the parameters $\Pran$, $\Le$ and $R_{\rho}$, and is of the order of $10 d$ 
for low $\Pran$ and high $\Le$ in planar domains \cite{Schmitt1983,Brown2013,Garaud2018}.
It is unaffected by rotation in unbounded gradient layers \cite{Sengupta2018}.

To quantify the evolution of the azimuthal length of the radial flow with $\Ra_t$,  we estimate the length $\ell$ from the peak of the poloidal kinetic energy spectra
plotted as a function of the order $m$ of the spherical harmonics.
The spectra averaged over the whole radial domain are shown in Figure \ref{fig:spectra_E5} for \mbox{$\Ra_t=-\Ra_c/3$} at \mbox{$\Ek=10^{-5}$}.
The spectra are taken from data snapshots but are representative of the dynamics.
For 
$\Ra_c\leq 1.2\times10^9$, the spectra have a well defined peak, which moves towards higher $m$ as $\Ra_c$ increases.
For $\Ra_c\geq 2\times10^9$, the spectra are relatively flat over a wide range of $m$, from $m=0$ up to $m$ greater than 100. 
This range widens towards larger $m$ as $\Ra_c$ increases. Fingering convection can therefore excite a wide range of lengths, from the largest scale to the finger scale.
However, there is no visible self-organisation of the fingers into large-scale clusters (see Figure~\ref{fig:equat_merid}) 
as in the high-$\Pran$ 3D planar simulations of \citeA{Paparella2012}.
At higher wavenumbers, the slope of the spectra is very steep for all $\Ra_c$. Power laws are indicated in Figure~\ref{fig:spectra_E5} for guidance. 
The spectra steepens in the high wavenumber range when $\Ra_c$ increases, being close to a power law with exponent $-5$ at the smallest $\Ra_c$
to $-10$ at the largest $\Ra_c$. 
In this range, the dynamics is dominated by viscous processes.  The steep slopes of the kinetic energy spectra at lengthscales smaller than the finger scale have been
observed in previous simulations of fingering simulations (\eg  \citeA{Paparella2012,Xie2017}).

\begin{figure}[h]
	\centering
	\includegraphics[clip=true,width=14cm]{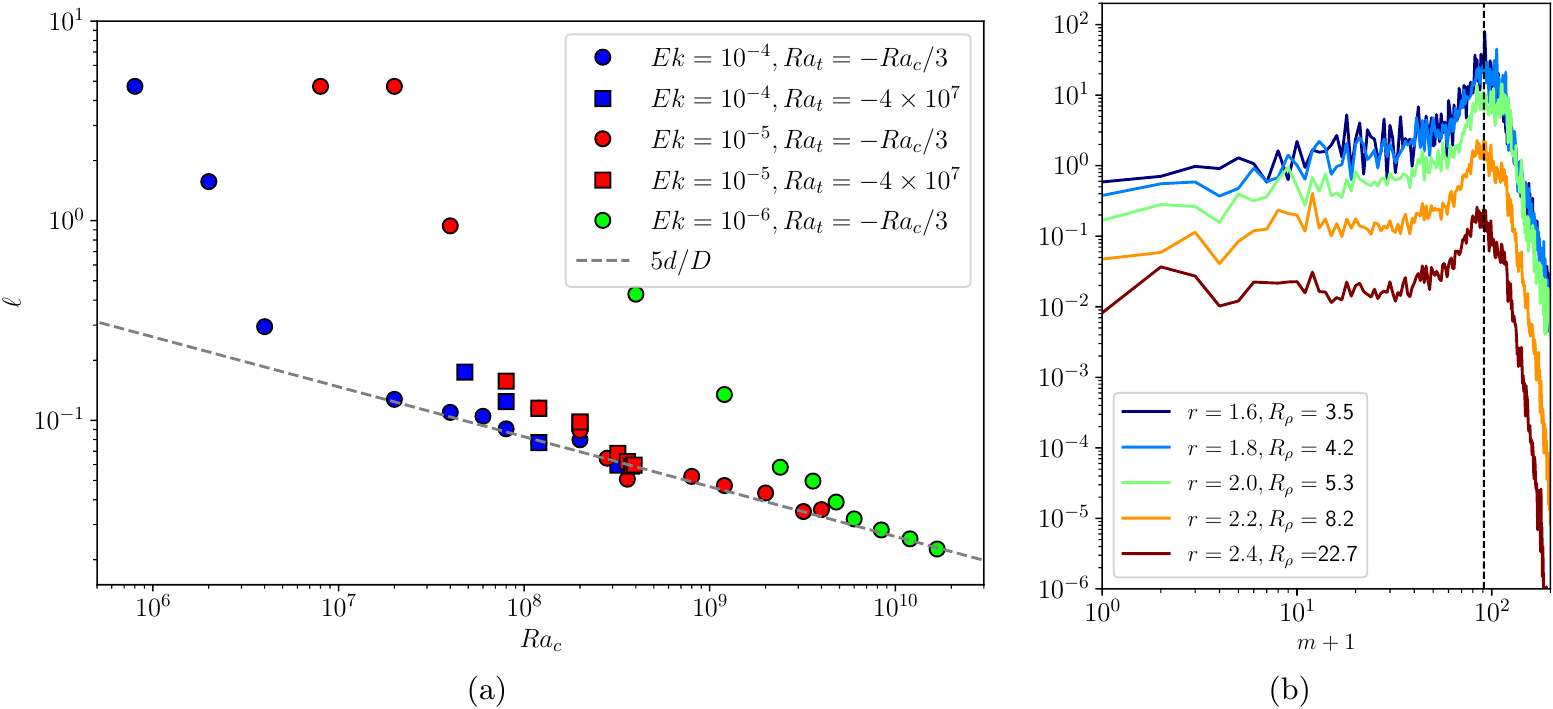}
	\caption{(a) Dominant azimuthal length of the poloidal flow, $\ell$, as a function of $\Ra_c$ for \mbox{$\Ra_t=-\Ra_c/3$} (circles) and 
	\mbox{$\Ra_t=-4\times10^7$} (squares).
	The theoretical scaling \mbox{$\ell=5d/D=5(r_o/|\Ra_t|)^{1/4}$} is plotted for the case \mbox{$\Ra_t=-\Ra_c/3$}.
	(b) Spectra of the poloidal kinetic energy (snapshot) as a function of the azimuthal order $m$ at different radii for \mbox{$\Ek=10^{-5}$}, 
	\mbox{$\Ra_c=8\times10^8$}
	and \mbox{$\Ra_t=-\Ra_c/3$}. The vertical dashed line represents the location of the peak $m_p$ determined from the radially-averaged spectrum.}
	\label{fig:length}
\end{figure}

The dominant azimuthal length of the poloidal flow, $\ell$, is defined as $\ell=\pi r_i /m_{p}$, where $m_p$ corresponds to the peak identified on the spectra 
of the poloidal kinetic energy (excluding the axisymmetric mode $m=0$). 
At the largest Rayleigh numbers, where the spectra is flat at small wavenumbers and the spectral peak is not clearly defined (for $m>0$), 
$m_p$ is selected to be the approximate value of $m$ before the drop-off at high wavenumbers.
Figure~\ref{fig:length}a shows $\ell$ as a function of $\Ra_c$ for $\Ra_t=-\Ra_c/3$.
As previously discussed, for small Rayleigh numbers close to the onset of fingering convection ($\Ra_c \Ek\leq10^3$), the flow has a large azimuthal length. 
After the first few points (\ie the first point of the dataset $\Ek=10^{-4}$ and the first two points of the dataset $\Ek=10^{-5}$, which have $m_p=1$, \ie $\ell\approx5$), 
$\ell$ decreases rapidly and then follows a shallower slope close to the theoretical scaling $|\Ra_t|^{-1/4}$ for $\Ra_c \Ek>10^3$. 
Interestingly, in this range, the data points follow a similar power law 
irrespective of the Ekman number. The azimuthal length of fingering convection is therefore not strongly affected by the Ekman number. 
We find that $\ell=5 d/D$ is a good fit to our data for $\Ra_c \Ek>10^3$.
The existence of two regimes (one at small $\Ra_c$ characterised by a steep decrease of $\ell$ with $\Ra_c$
and one at larger $\Ra_c$ characterised by the non-rotating scaling $\ell\sim|\Ra_t|^{-1/4}$) is also observed in \citeA{Monville2019}. 
They identify that the transition between the two regimes occurs at $N/\Omega \approx 0.5$ ($\Ra_c \approx \Ek^{-2}$ in our model for $\Ra_t=-\Ra_c/3$),
although this is based on simulations performed at one fixed value of the Ekman number ($\Ek=10^{-5}$).
In our simulations, the transition occurs earlier, especially for the smallest $\Ek$, and 
is better captured by $\Ra_c \Ek\approx10^3$.

There is no simple scaling law for the dependence of the horizontal length of the fastest growing mode on the density ratio $R_{\rho}$,
so this needs to be determined numerically \cite{Schmitt1983}.
To study how the azimuthal finger length varies with $R_{\rho}$ in our simulations, 
Figure~\ref{fig:length}a shows $\ell$ as a function of $\Ra_c$ for 
$\Ra_t=-4\times 10^7$ and $\Ek=\{10^{-5},10^{-4}\}$. 
The first point for each dataset is located at the edge of the stability domain, close to $R_{\rho}(r_i)=\Le$, in which case large scale modes 
are preferred as seen in \citeA{Monville2019}. As $\Ra_c$ increases (\ie as $R_{\rho}$ decreases),
$\ell$ decreases and the data points superpose fairly well with the case $\Ra_t=-\Ra_c/3$. 
The azimuthal finger length has therefore a small sensitivity to the density ratio and slowly decreases with it.
However $\Le=10$ in our study, so $\Ra_c$ (and $R_{\rho}(r_i)$) can only vary by a decade for fixed $\Ra_t$ and the variation in the azimuthal length cannot be extensively tested here.

The dominant azimuthal lengths were calculated from radially-averaged spectra. However, $N_t^2$ and $N_c^2$ vary with radius so this might produce 
radial variations of the flow length. 
This being said, $N_t^2$ only has modest variations across the layer 
($N_t^2(r_i)/N_t^2(r_o)=1/\asp=1.67$, see Figure~\ref{fig:N2}a), so the variation of the theoretical azimuthal length $d$ with radius is small 
($d(r_i)/d(r_o)=0.88$). However the variation of $R_{\rho}$ across the layer is much greater (see Figure~\ref{fig:N2}b).
To assess whether the azimuthal length of the poloidal flow changes with radius, Figure~\ref{fig:length}b shows the spectra of the poloidal kinetic 
energy at different radius for $\Ek=10^{-5}$, 
$\Ra_c=8\times10^8$ and $\Ra_t=-\Ra_c/3$. 
The spectra are similar at all radius, without any visible shift in the azimuthal wavenumber $m_p$ corresponding to the spectral peak.
The main difference is that spectra at larger radii (especially those for which $r>r_s$ where $r_s=2.26$) have a smaller amplitude than those located closer to the inner boundary.
Since the peak of the spectra does not shift with the radius, this implies that the dominant azimuthal length at different radii (which is proportional to $r/m_p$) 
increases linearly with radius. This increase of the azimuthal lengthscale with $R_{\rho}$ is consistent with the results discussed previously from Figure~\ref{fig:length}a.
Note that this linear increase of the dominant azimuthal length with radius is not particularly visible on the equatorial cross-section of Figure~\ref{fig:equat_merid} 
as the radial velocity extends to $r=r_s$ at most, corresponding to an increase of the azimuthal length of less than 50\%.

In planar simulations of rotating oscillatory double-diffusive convection, \citeA{Moll2017} found that double-diffusive convection is influenced by rotation when the 
modified Taylor number, $\Ta^{\ast}=4\Omega^2 d^4/\kappa_t^2$, is greater than unity. 
By definition of $d$ (equation~\eqref{eq:ltheo}),
$\Ta^{\ast}=4\Pran_t\Omega^2/N_t^2$, so we expect that the rotation will have an important effect on the flow for $N_t^2/\Omega^2<4\Pran_t\approx1$. 
Figure~\ref{fig:Le10} shows that 
all of our simulations are within the rotationally-dominated regime according to this condition. 
On the one hand, the effect of the rotation on the flow is visible on the 3D isosurfaces of Figure~\ref{fig:isosurf} as the fingers are elongated along the rotation axis
due to the Proudman-Taylor constraint, so their typical axial length is clearly affected by rotation. 
On the other hand, the azimuthal finger length is unaffected by rotation as it follows the non-rotating scaling law
and is independent of $\Ek$.

\subsection{Scaling of the radial velocity}

\begin{figure}[h]
	\centering
	\includegraphics[clip=true,width=\textwidth]{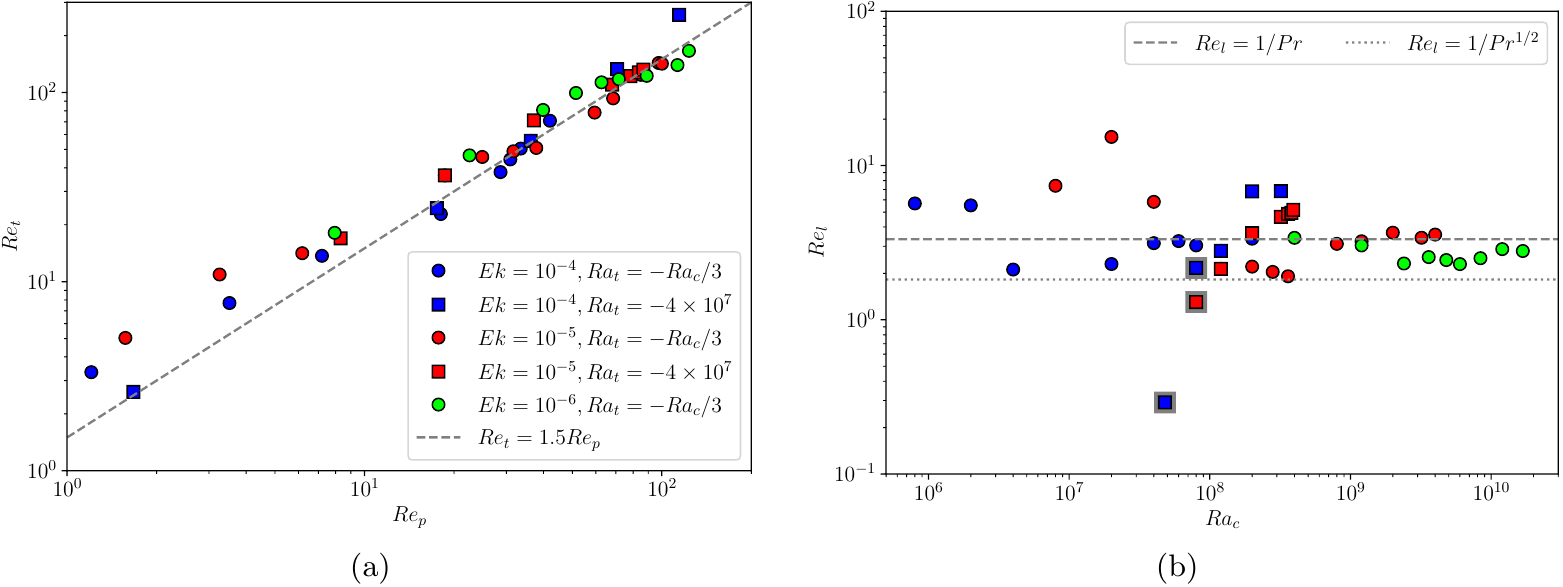}
	\caption{(a) Non-axisymmetric toroidal Reynolds number $\Rey_t$ as a function of the non-axisymmetric poloidal Reynolds number $\Rey_p$; 
	(b) Local Reynolds numbers $\Rey_{\ell}$ as a function of $\Ra_c$ for \mbox{$\Ra_t=-\Ra_c/3$} (circles) and
	\mbox{$\Ra_t=-4\times10^7$}  (squares). In (b), points with a thick grey edge correspond to cases with $r_{\rho}(r_i)>\Pran$.}
	\label{fig:Re}
\end{figure}

First, we assess the anisotropy between the radial and horizontal velocities of the non-axisymmetric flow (\ie corresponding to $m>0$).
To measure a typical radial velocity, we use the r.m.s. non-axisymmetric poloidal velocity $U_p$, which is calculated from time averaging 
the volumetric kinetic energy of the non-axisymmetric poloidal flow. Similarly, the horizontal velocity is approximated by the r.m.s. non-axisymmetric 
toroidal velocity $U_t$. The non-axisymmetric poloidal and toroidal Reynolds numbers are defined as
\begin{linenomath*}
\begin{equation}
	\Rey_p= \frac{U_p D}{\nu}, \quad \Rey_{t} = \frac{U_t D}{\nu}.
\end{equation}
\end{linenomath*}
Figure~\ref{fig:Re}a shows $\Rey_t$ as a function of $\Rey_p$. 
We find that a linear relation $\Rey_t= 1.5 \Rey_p$ is a good fit for the data point at large Rayleigh numbers. 
This is similar to the results of \citeA{Sengupta2018} and \citeA{Mather2021}.
The anisotropy between non-axisymmetric radial and horizontal velocities is therefore small and constant for varying Rayleigh numbers.

Second, we want to determine a scaling for the typical radial velocity as a function of the model parameters, which we could extrapolate to core conditions.
Fingering convection requires that the thermal diffusion acts on the horizontal finger scale $\ell$ on a timescale comparable with the advection timescale. 
We might therefore expect that the typical radial velocity of the fingers is of the order of $\kappa_t/\ell$ \cite{Radko2013}. 
This leads to the following theoretical scaling law for the local Reynolds number \mbox{$\Rey_{\ell}=\Rey_p \ell$},
\begin{linenomath*}
\begin{equation}
	\Rey_{\ell} = \frac{U_p \ell}{\nu} \sim \frac{1}{\Pran}.
	\label{eq:scaling1}
\end{equation}
\end{linenomath*}
However, for small Prandtl numbers, \citeA{Brown2013} and \citeA{Sengupta2018} argue that the vertical velocity of the fingers 
does not follow the scaling~\eqref{eq:scaling1} because the saturation of the fingering instability is caused by a secondary shear instability (Radko \& Smith 2012). 
In this case, the vertical velocity scales as $\lambda/k$, where $\lambda$ and $k$ are the growth rate and horizontal 
wavenumber of the fastest-growing linearly unstable mode. 
\citeA{Brown2013} obtain estimates for $\lambda$ and $k$ depending on the value of the reduced density ratio,
\begin{linenomath*}
\begin{equation}
	r_{\rho}=\frac{R_{\rho}-1}{\Le-1}.
\end{equation}
\end{linenomath*}
For $r_{\rho}\ll \Pran\ll1$ and $\Le\gg 1$, 
\citeA{Brown2013} shows that $\lambda\approx \sqrt{\Pran} \kappa_t/d^2$ and $k\approx1/d$. 
The theoretical scaling for the local Reynolds number is thus
\begin{linenomath*}
\begin{equation}
	\Rey_{\ell} \sim \frac{1}{\Pran^{1/2}},
	\label{eq:scaling2}
\end{equation}
\end{linenomath*}
where we assumed that $\ell\sim d/D$.
For larger density ratio, when $\Pran\ll r_{\rho}\ll1$,  \citeA{Brown2013} shows that the growth rate now also depends on $r_{\rho}$, $\lambda\approx \sqrt{\Pran/\Le r_{\rho}} \kappa_t/d^2$.
In this case, the local Reynolds number depends on both $R_{\rho}$ and $\Pran$,
\begin{linenomath*}
\begin{equation}
	\Rey_{\ell} \sim \frac{1}{((R_{\rho}-1)\Pran)^{1/2}}.
	\label{eq:scaling3}
\end{equation}
\end{linenomath*}

Figure~\ref{fig:Re}b shows the evolution of $\Rey_{\ell}$ with $\Ra_c$ for cases with $\Ra_t=-\Ra_c/3$ and $\Ra_t=-4\times10^7$. 
Most of our cases have $r_{\rho}(r_i) \lesssim \Pran$ (indeed $r_{\rho}(r_i)=0.26$ for $\Ra_t=-\Ra_c/3$), so we expect the local Reynolds number to be mostly constant.
This is indeed what we observe with $\Rey_{\ell}=\mathcal{O}(1)$ for the cases with $\Ra_t=-\Ra_c/3$, in agreement with the results of \citeA{Monville2019}. 
The theoretical scalings $1/\Pran$ and $1/\Pran^{1/2}$ are plotted for reference. Since our Prandtl number is not particularly small ($\Pran=0.3$), the difference between the two scalings is not large, so we cannot distinguish between the two. 
For 
$\Ra_t=-4\times 10^7$, we only have a few points in the parameter range $r_{\rho}(r_i)>\Pran$ (corresponding to $R_{\rho}(r_i)>3.7$ \ie $\Ra_c<1.08 \times10^8$),  
so verifying the theoretical scaling~\eqref{eq:scaling3} is not very practical. 
The first few points of the dataset  
$\Ra_t=-4\times 10^7$ correspond to this range, and for these we found that $\Rey_{\ell}$ decreases when $\Ra_c$ decreases (or equivalently,
when the density ratio increases) in qualitative agreement with the scaling~\eqref{eq:scaling3}.

\subsection{Zonal flow}

\subsubsection{Driving mechanism}

\begin{figure}[h]
	\centering
	\includegraphics[clip=true,width=\textwidth]{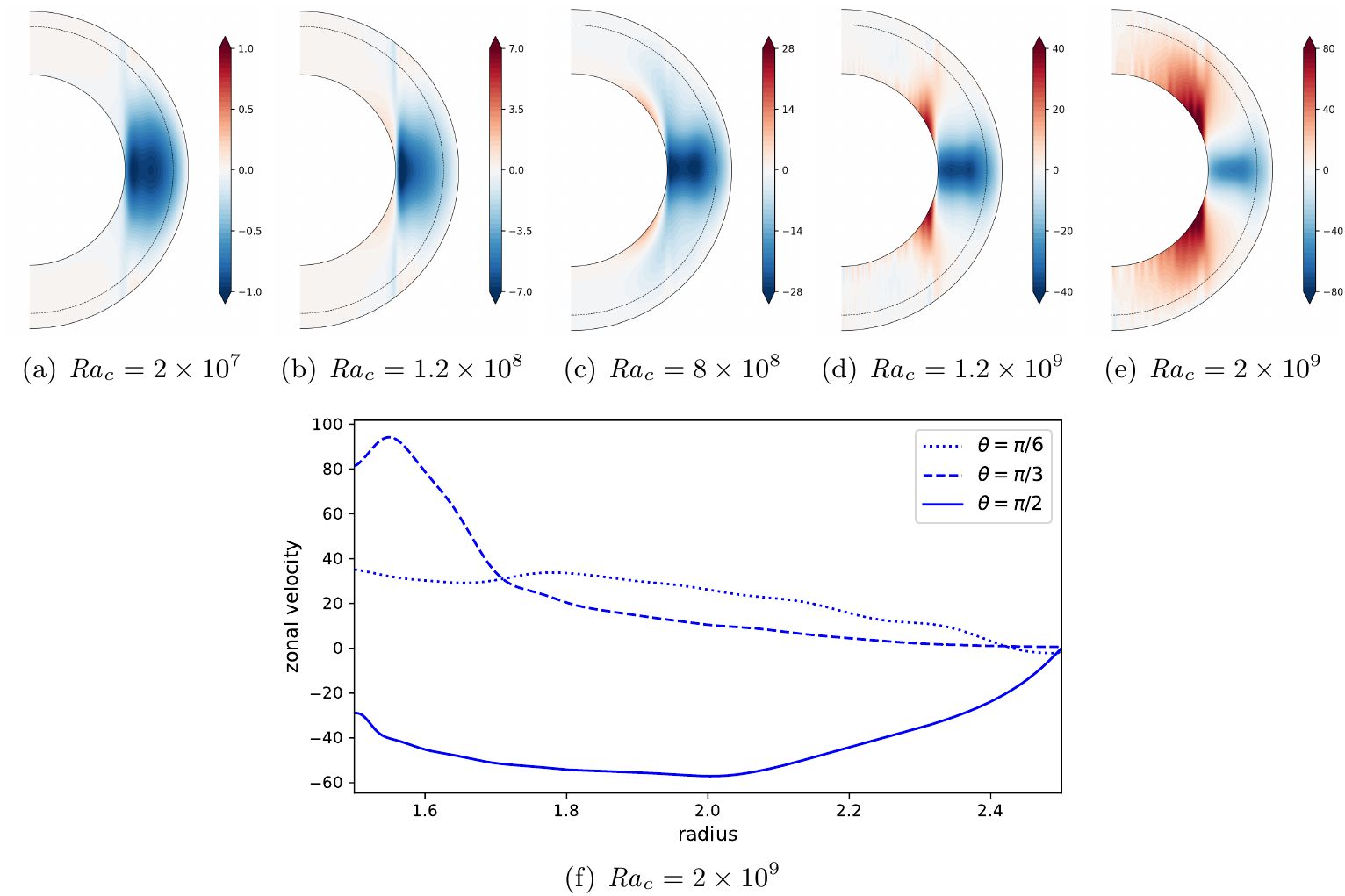}
	\caption{(a)-(e) Meridional cross-sections of the zonal velocity for \mbox{$\Ek=10^{-5}$} and \mbox{$\Ra_t=-\Ra_c/3$} (time averages).
	The dashed line represents \mbox{$r=r_s$}, the radius at which \mbox{$R_{\rho}=\Le$}.
	(f) Radial profile of the zonal velocity at different latitudes for $\Ra_c=2\times10^9$.}
	\label{fig:Uaxi_E5}
\end{figure}

The kinetic energy associated with the axisymmetric flow (\ie corresponding to $m=0$) grows as $\Ra_c$
increases, with $m=0$ becoming dominant over the non-axisymmetric modes at the largest Rayleigh numbers. 
Most of the axisymmetric energy is contained in the azimuthal component of the flow (\ie the zonal flow), so we turn our attention to the formation of this component. 
Figure~\ref{fig:Uaxi_E5} shows meridional cross-sections of the time-averaged zonal velocity for different Rayleigh numbers for $\Ek=10^{-5}$ and $\Ra_t=-\Ra_c/3$.
In all cases, the zonal flow is symmetric with respect to the equatorial plane and is retrograde in the equatorial region. As $\Ra_c$ increases, a prograde zonal flow appears
inside the tangent cylinder. 
The prograde velocity inside the tangent cylinder can be nearly twice larger than the retrograde velocity in the equatorial plane.
Radial profiles of the zonal velocity at different latitudes are plotted in Figure~\ref{fig:Uaxi_E5}(f) for one of the largest Rayleigh numbers.
The profiles show that the zonal flow penetrates deeply inside the outer region that is linearly stable to fingering convection ($r>r_s$ where $r_s=2.26$).
We observe no significant temporal variations of the zonal flow in all the simulations. 
Simulations performed in the case  
$\Ra_t=-4\times 10^7$, where the density ratio varies between $1$ and $\Le$, have similar zonal flows, where the
zonal velocity inside the tangent cylinder becomes progressively prograde as $R_{\rho}(r_i)$ gets close to $1$ (\ie as $\Ra_c$ increases).
 
In simulations of fingering convection in rotating spherical shells at small Rayleigh numbers ($\Ra_c\Ek<10^3$), \citeA{Mather2021} find a similar structure of the  
zonal flow, although their prograde flow is located outside the tangent cylinder, which might be a consequence of the wider gap geometry ($\asp=0.35$) used in their study.  
In a full sphere geometry, \citeA{Monville2019} found that equatorially-antisymmetric flows emerge on long time scales and becomes dominant over the equatorially-symmetric flows 
after several viscous diffusion times. Although a number of our simulations were performed over many viscous diffusion times 
(\eg more than 350 viscous timescales in the case shown in Figure~\ref{fig:t_KE_E5Ra5e7}), 
we did not observe the emergence of equatorially-antisymmetric flows, which might be a feature restricted to the full sphere geometry. 

\begin{figure}[h]
	\centering
	\includegraphics[clip=true,width=\textwidth]{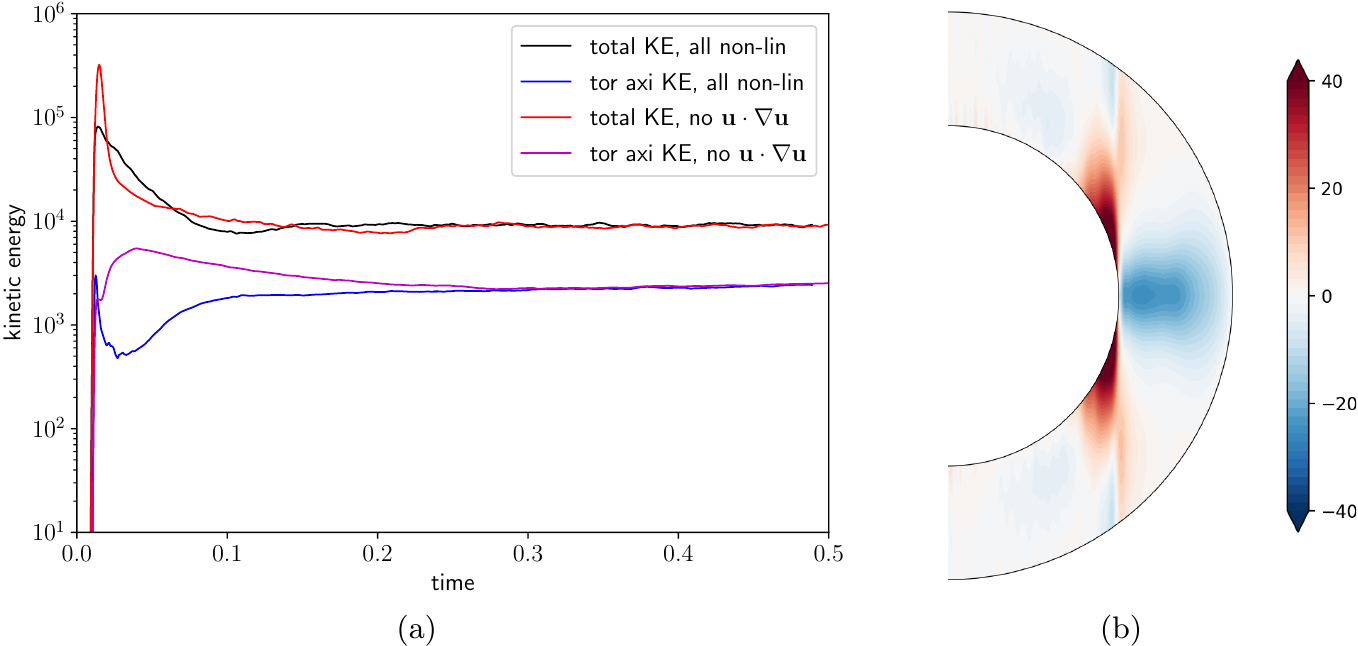}
	\caption{(a) Time series of the total kinetic energy and the toroidal axisymmetric kinetic energy in a fully non-linear simulation
	and in a simulation where the \mbox{$(\vel \cdot \nabla) \vel$} term is removed from the Navier-Stokes equations (for the same parameters:
	\mbox{$\Ek=10^{-5}$}, 
	\mbox{$\Ra_c=1.2\times10^9$} and $\Ra_t=-\Ra_c/3$). (b) Meridional cross-sections of the zonal velocity (time averaged) in the 
	simulation with no \mbox{$(\vel \cdot \nabla) \vel$} term.}
	\label{fig:nougu}
\end{figure}

The evolution equation for the zonal velocity is given by
\begin{linenomath*}
\begin{equation}
	\pd{\zw}{t} = - \overline{[(\vel \cdot \vn) \vel]_{\phi} } - \frac{2}{\Ek} \overline{u_s} + \overline{[\vn^2 \vel]_{\phi}},
	\label{eq:zw}
\end{equation}
\end{linenomath*}
where the overbar denotes an azimuthal average. The zonal flow is driven either by the Reynolds stress (corresponding to the first term on RHS) or 
by the Coriolis force that deflects the meridional circulation (second term on the RHS), while the viscous term (third term on the RHS) corresponds to a sink term.
To assess the role of the Reynolds stress, we run a simulation at 
$\Ra_c=1.2\times10^9$, where we artificially remove the \mbox{$(\vel \cdot \nabla) \vel$} term 
in the Navier-Stokes equation (while keeping the non-linear terms in the temperature and composition equations). The time series of the total kinetic energy 
and the toroidal axisymmetric kinetic energy (\ie the energy corresponding to the zonal flow) are shown in Figure~\ref{fig:nougu}a and compared with the time series from the fully nonlinear simulation. 
The energies in the ``altered" simulation saturate at a similar level than the energies of the fully nonlinear simulation, 
meaning that the system mainly saturates due to the non-linearities in the composition equation. 
(An additional altered simulation confirms that the simulation does not saturate when removing only the term \mbox{$\vel \cdot \vn \Cpert$} in the composition equation.)
The meridional cross-section of the time-averaged zonal velocity of the altered simulation (without the \mbox{$(\vel \cdot \nabla) \vel$} term)
is shown in Figure~\ref{fig:nougu}b and can be compared with Figure~\ref{fig:Uaxi_E5}d. 
The zonal flow of the altered simulation is remarkably similar to the zonal flow of the fully nonlinear simulation, and so, we conclude that the main source of the zonal flow 
is not the Reynolds stress but the Coriolis force acting on the meridional circulation.

\begin{figure}[h]
	\centering
	\includegraphics[clip=true,width=10cm]{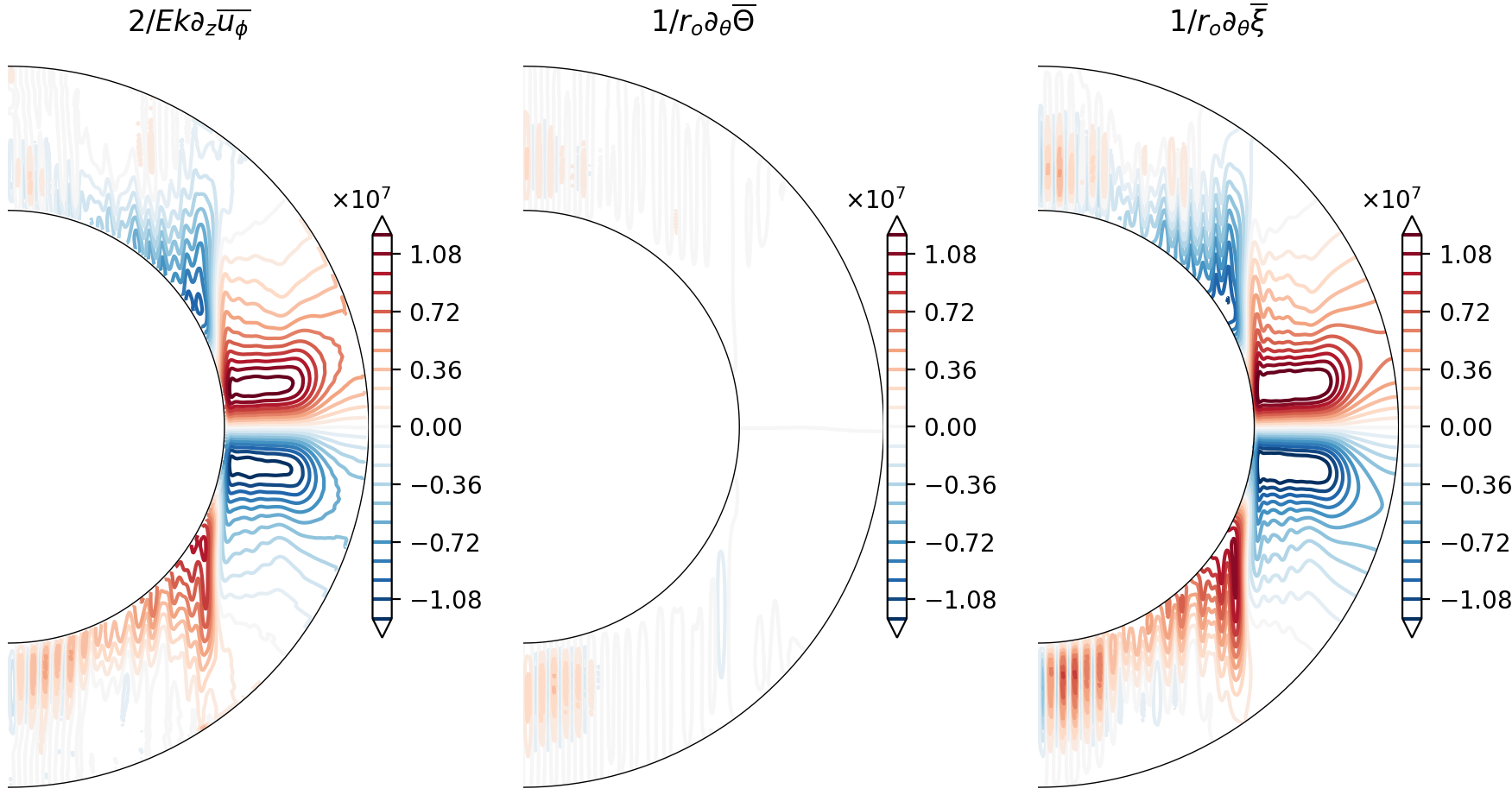}
	\caption{Terms in the thermal wind equation for \mbox{$\Ek=10^{-5}$}, 
	\mbox{$\Ra_c=1.2\times10^9$} and \mbox{$\Ra_t=-\Ra_c/3$}.}
	\label{fig:TW_E5}
\end{figure}

\begin{figure}[h]
	\centering
	\includegraphics[clip=true,width=\textwidth]{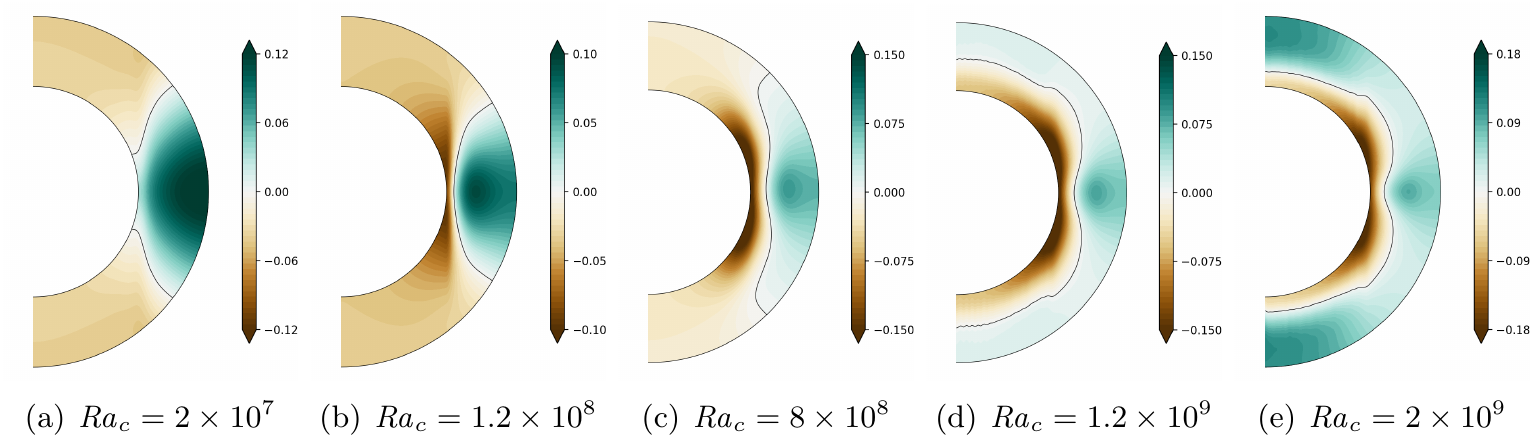}
	\caption{Meridional cross-sections of the axisymmetric composition perturbation for \mbox{$\Ek=10^{-5}$} and \mbox{$\Ra_t=-\Ra_c/3$}. 
	The composition is normalised by \mbox{$C_s(r=r_i)$} and time-averaged. The black line corresponds to the isoline $\overline{\Cpert}=0$.}
	\label{fig:Caxi_E5}
\end{figure}

To understand the mechanism driving the meridional circulation involved in Equation~\eqref{eq:zw}, we look at the evolution equation 
of the $\phi$-component of the axisymmetric vorticity, $\overline{\omega}_{\phi} = \partial \overline{u_s}/\partial z - \partial \overline{u_z}/\partial s$  
(as taking the curl of the Navier-Stokes equation has the advantage of eliminating the pressure gradient).
Neglecting the non-linear inertial terms, we get
\begin{linenomath*}
\begin{equation}
	\pd{\overline{\omega}_{\phi}}{t} = \frac{2}{\Ek} \frac{\partial \zw}{\partial z} - \frac{1}{r_o} \pl \frac{\partial \overline{\Tpert}}{\partial \theta} +\frac{\partial \overline{\Cpert}}{\partial \theta} \pr + \overline{[\vn^2 \boldsymbol{\omega}]_{\phi}}.
\end{equation}
\end{linenomath*}
In a steady state, and neglecting the viscous term, this equation is often called the thermal wind equation and shows that latitudinal variations of the axisymmetric composition or 
temperature perturbations are balanced by axial variations of the zonal velocity.
In other words, the latitudinal variations of density create a source of azimuthal vorticity; the associated meridional velocity is then deflected by the Coriolis force 
into a (z-dependent) zonal velocity. 
Figure~\ref{fig:TW_E5} shows meridional cross-sections of the terms in the thermal wind equation for the case 
$\Ra_c=1.2\times10^9$. 
The latitudinal variations of $\overline{\Cpert}$ mostly balances the $\partial \zw/\partial z$ term everywhere in the bulk, with the latitudinal variations of $\overline{\Tpert}$ playing a small
opposing role. The zonal flow is therefore indirectly due to the latitudinal variations of the axisymmetric composition. 
Figure~\ref{fig:Caxi_E5} shows the axisymmetric composition perturbation for the same simulations as in Figure~\ref{fig:Uaxi_E5}.
 At small Rayleigh numbers where the fingering convection only occurs outside of the tangent cylinder 
($\Ra_c<8\times10^8$), a positive $\overline{\Cpert}$ accumulates in the equatorial region near the outer boundary. 
Regions at higher latitudes and nearer to the inner boundary are then depleted in light elements, which creates a positive latitudinal gradient of composition in the northern 
hemisphere. 
When $\Ra_c$ increases and the fingering convection develops in the polar regions, the polar regions now becomes enriched in light elements at large radius, while the mid-latitude are 
comparatively more depleted than poles and equator, which creates a latitudinal gradient of composition that changes sign at mid-latitudes.
Accordingly, the zonal flow (which must vanish at $r=r_o$) is prograde inside the tangent cylinder and retrograde outside.

\subsubsection{Scaling of the zonal velocity}
\label{sec:scalinguzonal}

\begin{figure}[h]
	\centering
	\includegraphics[clip=true,width=7.5cm]{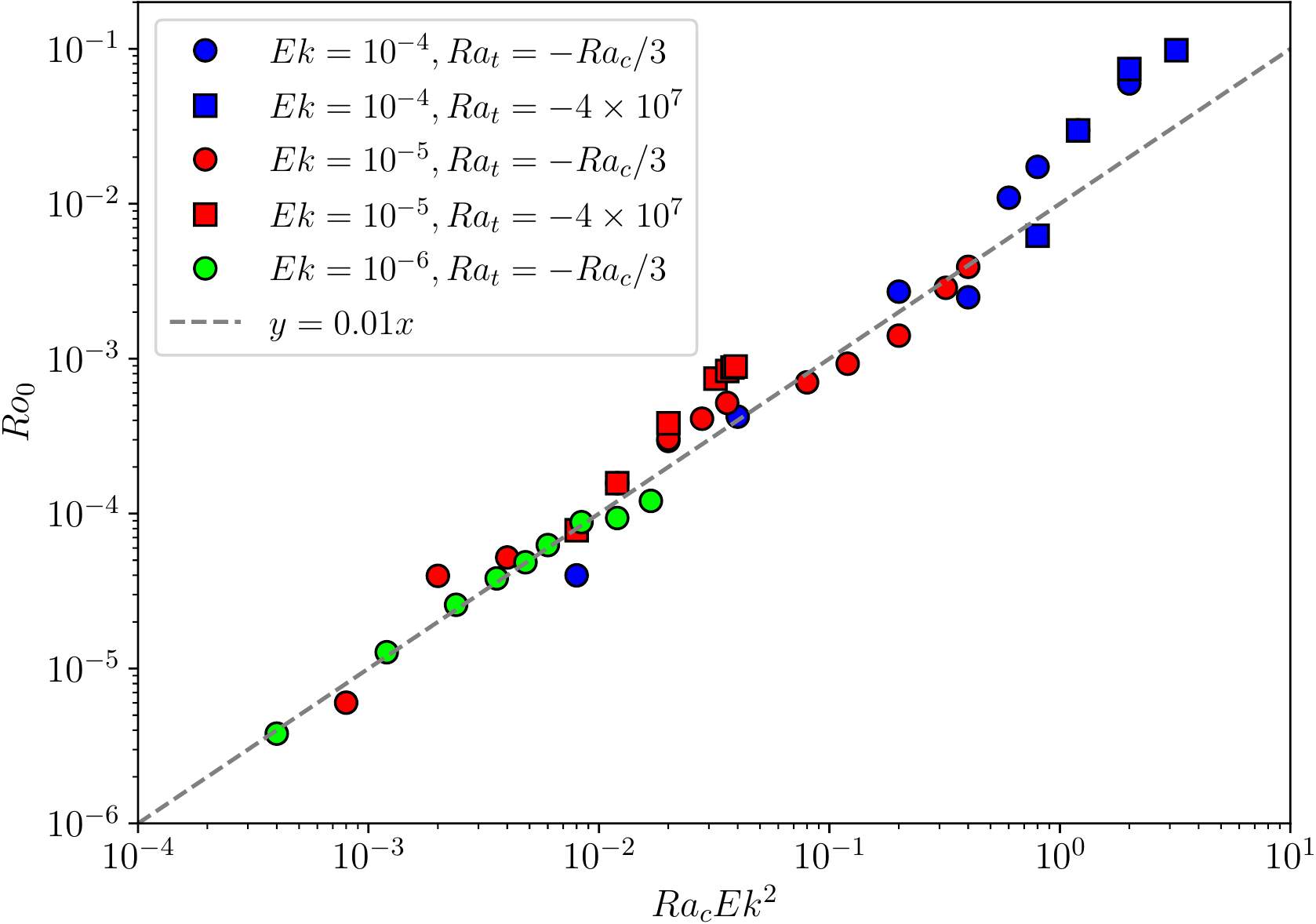}
	\caption{Rossby number based on the zonal flow amplitude, $\Ro_0$, as a function of \mbox{$\Ra_c \Ek^2$}.}
	\label{fig:Ro0}
\end{figure}

To obtain an approximate scaling of the zonal flow amplitude with the model parameters, we use an order of magnitude estimate from the thermal wind equation
\begin{linenomath*}
\begin{equation}
	\frac{2}{\Ek} \frac{\Rey_0}{H_z} \sim \frac{1}{r_o} \frac{\Delta \Cpert}{H_{\theta}},
\end{equation}
\end{linenomath*}
where $\Rey_0=U_0 D/\nu$ with $U_0$ the r.m.s. value of the zonal flow and $\Delta \Cpert>0$ is an estimate of the variation of the composition perturbation across the domain.
We will assume that the axial and latitudinal variations occur over lengths $H_z\sim H_{\theta} \sim 1$.
Since the diffusion of composition is slow, we assume that $\Delta \Cpert$ is of the same order as the difference of the background composition 
over a mixing length $H_r$.
This implies that
\begin{linenomath*}
\begin{equation}
	\frac{\Delta \Cpert}{H_r} \sim \left| \frac{\textrm{d} C_s}{\textrm{d}r} \right| \sim \frac{1}{r_i} \frac{Ra_c}{\Sc}.
\end{equation}
\end{linenomath*}
As a result, we obtain that the Rossby number based on the zonal flow, $\Ro_0=U_0/\Omega D$, scales as
\begin{linenomath*}
\begin{equation}
	\Ro_0 = \Rey_0 \Ek \sim \frac{H_r \Ra_c \Ek^2}{2 r_o r_i \Sc}.
	\label{eq:scalingRo}
\end{equation}
\end{linenomath*}
The radial flows extend throughout most of the layer depth (see Figure~\ref{fig:equat_merid}),
so, as a first approximation, we assume that the mixing length is constant and is close to the layer depth, $H_r=\mathcal{O}(1)$.
We therefore expect $\Ro_0$ to depend linearly on $\Ra_c \Ek^2$. 
Figure~\ref{fig:Ro0} shows $\Ro_0$ as a function of this parameter. 
The zonal Rossby number reaches values up to $0.1$, \ie the differential rotation associated with the zonal flow can be as large as 10\% of the background rotation.
The points from the datasets with different Ekman numbers superpose fairly well and they approximately follow a linear scaling, 
in agreement with the theoretical scaling ~\eqref{eq:scalingRo}.
The theoretical scaling has a factor $1/2r_ir_o\Sc\approx0.04$, while we find an empirical factor 
that is approximately 
4 times smaller.
This suggests that the mixing length is approximately $H_r\approx0.25$. In section~\ref{sec:Nu} on the compositional transport,
we will study the validity of the assumption of constant $H_r$.

\subsubsection{Effect of the boundary conditions}

Since the zonal flows are mainly due to  latitudinal variations of the composition in our simulations, it is important to consider whether the
choice of boundary conditions for $\Tpert$ and $\Cpert$ influences the results.
So far we have used zero flux boundary conditions (equation~\eqref{eq:BC_TC}), 
motivated by the idea that the fluxes are fixed by the backgound state at both boundaries (and in particular there is no flux of light elements at the core-mantle boundary). 
We have performed  a small number of additional simulations with fixed $\Tpert$ and $\Cpert$ boundary conditions at $r=r_i$ and $r=r_o$:
\begin{linenomath*}
\begin{eqnarray}
	\Tpert(r=r_i) = \Tpert(r=r_o) = \Cpert (r=r_i) =  \Cpert (r=r_o) = 0.
\label{eq:newBC}
\end{eqnarray}
\end{linenomath*}
With this choice of boundary conditions, fluxes of temperature or composition perturbations are thus allowed in or out of the domain. 

\begin{figure}[h]
	\centering
	\includegraphics[clip=true,width=0.8\textwidth]{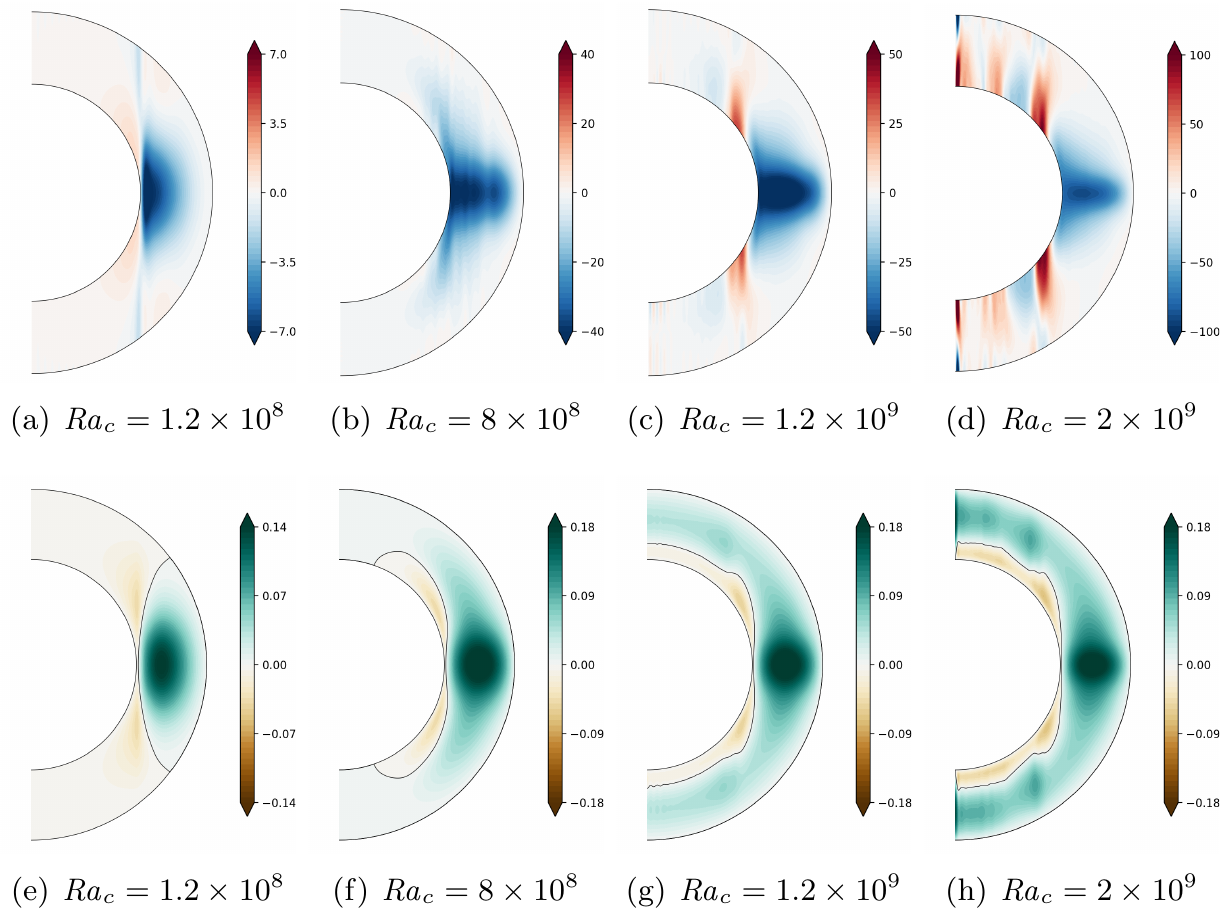}
	\caption{Meridional cross-sections of (a)-(d) the zonal velocity 
	and (e)-(h) the axisymmetric composition perturbation for \mbox{$\Ek=10^{-5}$} and \mbox{$\Ra_t=-\Ra_c/3$} (time averages) with 
	fixed $\Tpert$ and $\Cpert$ boundary conditions (equation~\eqref{eq:newBC}).
	Both fields are time-averaged.
	The composition is normalised by \mbox{$C_s(r=r_i)$} and the black line corresponds to the isoline $\overline{\Cpert}=0$.}
	\label{fig:UCaxi_E5_FT}
\end{figure}

Figure~\ref{fig:UCaxi_E5_FT} shows meridional cross-sections of the zonal velocity for the same parameters as in Figure~\ref{fig:Uaxi_E5} and using the
boundary conditions~\eqref{eq:newBC}. The case $\Ra_c=2\times10^7$ is not shown because fingering convection decays in this case. 
Similarly to the case with zero flux boundary conditions, the zonal flow is retrograde outside the tangent cylinder. 
The amplitude of the zonal flow is similar with both types of boundary conditions. 
The same observation is made in simulations of overturning thermal convection when the thermal boundary conditions are changed \cite{Clarte2021}.
In our simulations, the most notable difference is that 
zonal flows of alternating direction form close to the inner boundary inside the tangent cylinder at large Rayleigh numbers for fixed $\Tpert$ and $\Cpert$ boundary conditions.
The meridional cross-sections of the axisymmetric composition perturbation are also shown in Figure~\ref{fig:UCaxi_E5_FT} and can be compared with Figure~\ref{fig:Caxi_E5}.  
At the largest Rayleigh numbers, the accumulation of light elements at large radius inside the tangent cylinder is more patchy for fixed $\Tpert$ and $\Cpert$ boundary conditions.
The alternating zonal jet pattern is thus associated with rapid latitudinal variations of $\overline{\Cpert}$.
At 
$\Ra_c=2\times10^9$, the fields shown in Figure~\ref{fig:UCaxi_E5_FT} have been time-averaged over approximately $0.2$ viscous timescale, 
or equivalently $50$ zonal advection timescales (where one zonal advection timescale corresponds to $D/U_0$). 
The alternating zonal jets and the patchiness of the composition are robust features on this relatively short timescale, but it is plausible that they will evolve on longer time scales.
Overall the zero flux boundary conditions lead to a smoother latitudinal distribution of $\overline{\Cpert}$, hence the absence of multiple zonal jets inside the tangent cylinder.

\subsection{Convective transport}
\label{sec:Nu}

\begin{figure}[h]
	\centering
	\includegraphics[clip=true,width=\textwidth]{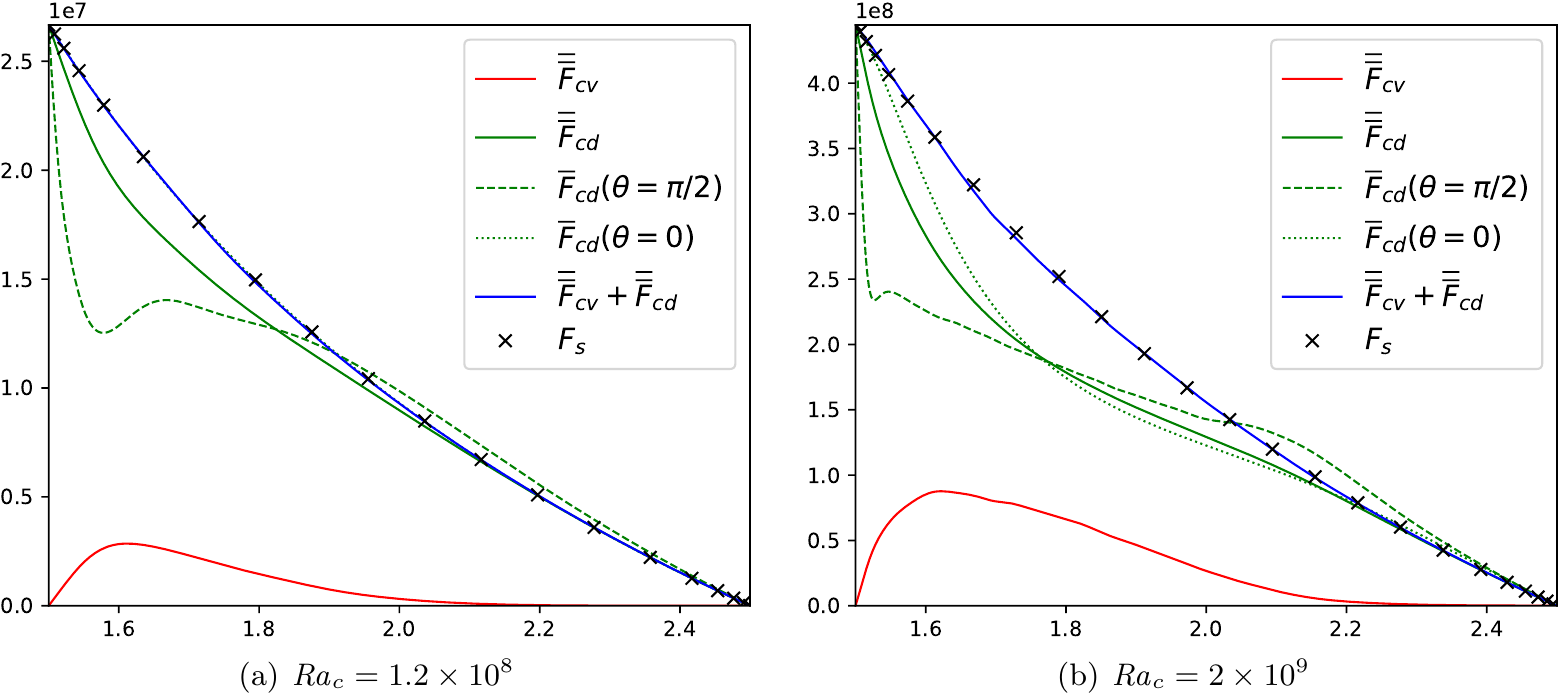}
	\caption{Radial profiles of the convective  and conductive fluxes of composition (\mbox{$F_{cv}$} and \mbox{$F_{cd}$} respectively) 
	for \mbox{$\Ek=10^{-5}$} and \mbox{$\Ra_t=-\Ra_c/3$}. 
	The single overbar denotes an azimuthal average and the double overbar a spherical average. $F_s$ is conductive flux in the absence of motion.
	The fluxes have been extracted from data snapshots and are not time-averaged. The snapshots are representative of the dynamics in each case.}
	\label{fig:flux}
\end{figure}

We now study the efficiency of the convective transport. We focus on the transport of composition as the convective transport
of heat is negligible in fingering convection \cite{Garaud2018}.
The convective flux of composition averaged over a spherical surface $\mathcal{S}(r)$ can be expressed as
\begin{linenomath*}
\begin{equation}
	\overline{\overline{F}}_{cv}(r) =  \Sc \int_{\mathcal{S}(r)} u_r \Cpert \mathrm{d}\mathcal{S},
\end{equation}
\end{linenomath*}
where the double overbar denotes a spherical average. The spherically-averaged conductive flux of composition is
\begin{linenomath*}
\begin{equation}
	\overline{\overline{F}}_{cd}(r) = - \int_{\mathcal{S}(r)} \frac{\partial \Cpert}{\partial r} \mathrm{d}\mathcal{S} + F_s (r),
\end{equation}
\end{linenomath*}
where $F_s(r)=-\textrm{d} C_s/\textrm{d}r$ is the static flux. 
In a steady state, we must have $\overline{\overline{F}}_{cd}(r)+\overline{\overline{F}}_{cv}(r) = F_s (r)$ at each radius.
Figure~\ref{fig:flux} shows the radial profiles of the fluxes for two different $\Ra_c$ at $\Ek=10^{-5}$ and $\Ra_t=-\Ra_c/3$. 
Although, the fluxes are calculated from data snapshots and are not time-averaged, the total flux $\overline{\overline{F}}_{cd}+\overline{\overline{F}}_{cv}$
is everywhere close to the static flux in both cases.  
The convective flux is smaller than the conductive flux everywhere in the volume for all $\Ra_c$ studied here. 
At the largest $\Ra_c$, the convective flux becomes significant over most of the radius $r<r_s$, but is completely negligible for $r>r_s$. 
The gradient of composition visibly weakens (in absolute value) in the fluid interior when $\Ra_c$ increases. 
Unlike in standard overturning convection, this observation is not necessarily expected. 
Indeed, as discussed previously, the local Reynolds number remains constant and the azimuthal lengthscale of the flow decreases 
with $\Ra_c$. One might therefore have expected the mixing efficiency to remain constant or to decrease. Clearly, this is not the case here. 
The conductive fluxes in the polar and equatorial regions (averaged in azimuth) are also plotted in Figure~\ref{fig:flux} . 
At the smallest $\Ra_c$, the radial profile of $\overline{F}_{cd}$ at the North pole superposes with $F_s$ because there is
no convection occurring inside the tangent cylinder. 
At the largest $\Ra_c$, the polar regions are now convecting but mixing at small radius ($r<1.8$) is visibly less efficient in the polar regions
than in the equatorial region.

\begin{figure}[h]
	\centering
	\includegraphics[clip=true,width=\textwidth]{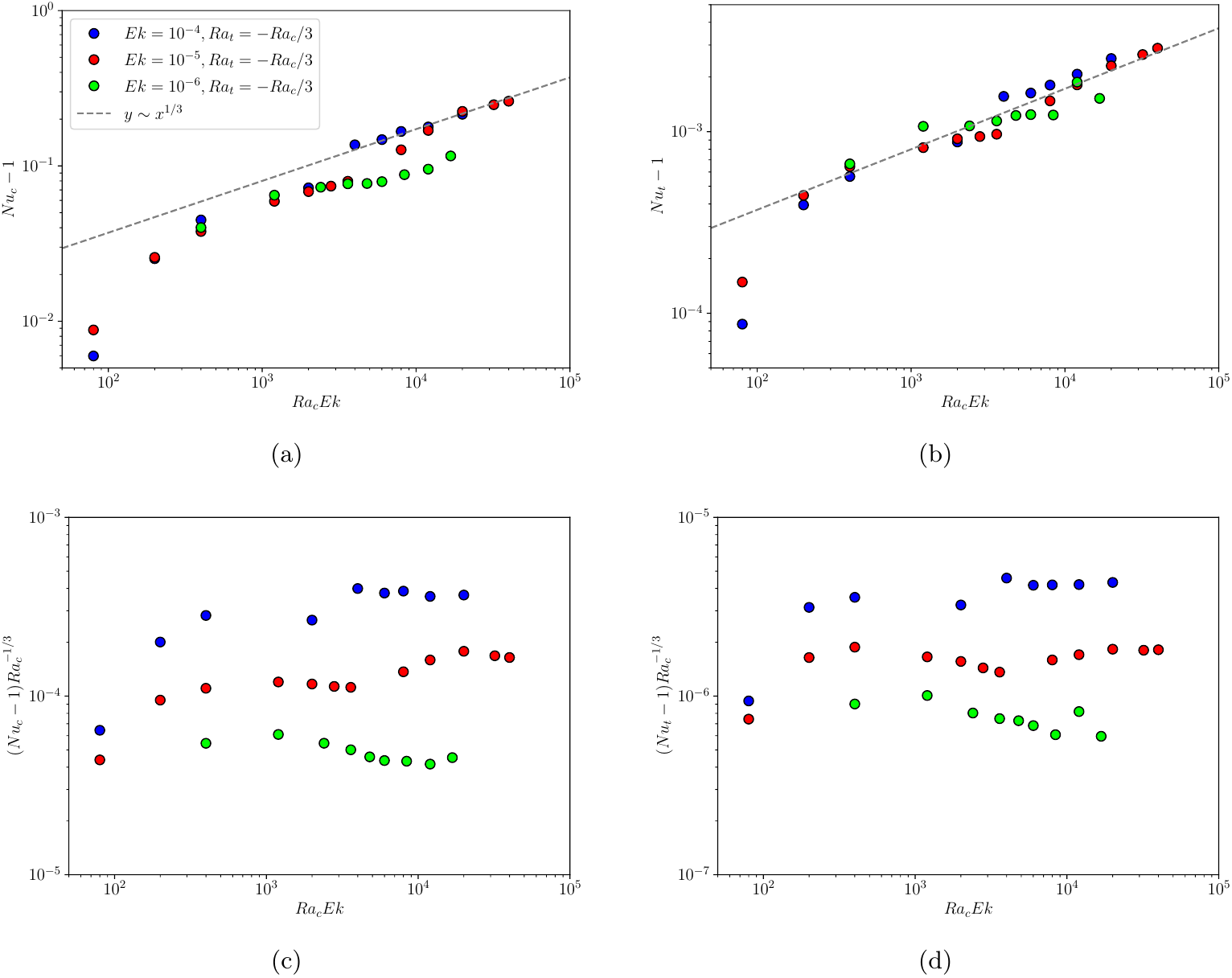}
	\caption{(a) Compositional and (b) thermal Nusselt numbers as a function of \mbox{$\Ra_c\Ek$} for \mbox{$\Ra_t=-\Ra_c/3$}.
	(c)-(d) Same plots as (a)-(b) for the compensated Nusselt numbers.}
	\label{fig:Nu}
\end{figure}

The efficiency of the convective transport is often quantified by a Nusselt number. The Nusselt number measures the ratio of the total flux
to the conductive flux and is by definition greater than one. For a system with fixed flux boundary conditions, the Nusselt number can be quantified from the difference of
spherically-averaged field across the layer \cite{Mound2017,Monville2019}. 
We therefore define the Nusselt number for the composition as 
\begin{linenomath*}
\begin{equation}
	\Nu_c = \frac{C_s(r_o)-C_s(r_i)}{C_s(r_o)-C_s(r_i)+\overline{\overline{\Cpert}}(r_o)-\overline{\overline{\Cpert}}(r_i)},
\end{equation}
\end{linenomath*}
and for the temperature
\begin{linenomath*}
\begin{equation}
	\Nu_t = \frac{T_s(r_o)-T_s(r_i)}{T_s(r_o)-T_s(r_i)+\overline{\overline{\Tpert}}(r_o)-\overline{\overline{\Tpert}}(r_i)}.
\end{equation}
\end{linenomath*}
Figure~\ref{fig:Nu} shows the variations of $\Nu_c$ and $\Nu_t$ as a function of $\Ra_c \Ek$. This combined parameter is chosen for the abscissa as 
it collapses fairly well the data points from the different series in $\Ek$. 
Values of $\Nu_t$ are always much smaller than $\Nu_c$, as expected for fingering convection at low $\Pran$ \cite{Garaud2018}. 
Both Nusselt numbers slowly increase with $\Ra_c$.
$\Nu_c-1$ never exceeds unity in our simulations.
No single power law can adequately fit all our data, but for $\Ra_c\Ek>10^3$, both Nusselt numbers appear to follow approximately the scaling relation 
$\Nu_{c,t}-1=C \Ra_c^{1/3}$ (see the compensated plots (c) and (d)), where the coefficient $C$ depends on $\Ek$. 
This scaling law (where $C$ depends more generally on $R_{\rho}$, $\Pran$ and $\Le$)
has been proposed for non-rotating planar fingering convection \cite{Turner1967} and verified 
in experiments and numerical simulations for $\Pran>1$ \cite{Taylor1996,Radko2000}.
On the contrary, for non-rotating fingering convection with $\Pran<1$, \citeA{Brown2013} propose a scaling law for $\Nu_c$ 
that is constant for fixed $R_{\rho}$, $\Pran$ and $\Le$. 
\citeA{Monville2019} identified two regimes for the evolution of the Nusselt numbers with $\Ra_c$, with a transition occurring at $N/\Omega\approx 0.5$:
for $N/\Omega<0.5$, the Nusselt numbers increase with $\Ra_c$ (but without following a single power law) and for $N/\Omega>0.5$, the Nusselt numbers become constant. 
Our parameter survey does not sample the domain $N/\Omega>0.5$ (corresponding to $\Ra_c>\Ek^{-2}$ in our simulations), 
so it is plausible that larger $\Ra_c$ are required to recover the constant scaling relation of \citeA{Brown2013} in our model.

To conclude this section, we return to the mixing length $H_r$ introduced in section~\ref{sec:scalinguzonal}. $H_r$ can be related to $\Nu_c$ as
\begin{linenomath*}
\begin{equation}
	\Nu_c = \frac{1}{1-H_r},
	\label{eq:Hr}
\end{equation}
\end{linenomath*}
with $H_r\leq 1$ by definition.
For small $H_r$, $H_r\approx \Nu_c-1$. 
Consequently, the mixing length increases when the Rayleigh numbers increase, despite the decrease of the azimuthal finger length. 
The approximate dependence of $H_r$ on $\Ra_c^{1/3}$ deduced from Figure~\ref{fig:Nu} at large Rayleigh numbers might explain why the data points in 
the scaling of the zonal flow amplitude in Figure~\ref{fig:Ro0} follow a slightly steeper power law than predicted by equation~\eqref{eq:scalingRo} when assuming constant $H_r$.
Figure S1 given in the Supporting Information shows $Ro_0$ as a function of the scaling~\eqref{eq:scalingRo} using $H_r$ deduced from Equation~\ref{eq:Hr}
and our numerical data.
The agreement between the data and the theoretical scaling is improved, especially at large values of $\Ra_c\Ek^2$, but overall
the relatively weak dependence of $H_r$ on $\Ra_c$ means that considering that $H_r$ is constant in Equation~\eqref{eq:scalingRo} is an acceptable assumption.
In particular, this assumption allows to predict the amplitude of the zonal flows as a function of input parameters only.

\section{Implications for planetary magnetic fields}
\label{sec:discussion}

\subsection{Dynamo action}

Using the scaling laws for the azimuthal finger length and the radial velocity, we can speculate on the interaction of fingering convection
with magnetic fields. 
First, we assess whether fingering convection might be able to sustain dynamo action. 
Indeed, if the stable layer could generate its own dynamo, its magnetic field could potentially cancel out some of the magnetic field
produced by the deep convective layer, leading to the weak magnetic field observed at the surface. 
The idea of two distinct dynamo regions in Mercury has been previously considered by \citeA{Vilim2010} with a model consisting of two nested convective layers
that could reproduce Mercury's observed field strength.
In dynamo simulations of fingering convection, \citeA{Mather2021} found that fingering convection
could not produce dynamo action, even at large magnetic Prandtl numbers, \ie small magnetic diffusivity ($\Pm=\nu/\eta$ with $\eta$ the magnetic diffusivity).
However, their calculations are performed at small Rayleigh numbers and are dominated by large-scale modes. 
The small-scale fingers obtained at larger Rayleigh numbers might have different dynamo properties. 
Indeed, as discussed in Section~\ref{sec:3D}, fingering convection has different kinetic helicity distributions at small and large Rayleigh numbers.
At large Rayleigh numbers, we find that the helicity distribution is close to the one obtained in rotating overturning convection, which has the ability
to produce large-scale magnetic fields \cite{Olson1999}, suggesting that small-scale fingers have the right ingredients for dynamo action.
Here we want to estimate whether the fingers might be able to generate a dynamo by using a minimal requirement based on the 
magnetic Reynolds number. 
The magnetic Reynolds number, $\Rm$, estimates the ratio of the magnetic diffusion timescale of a large-scale magnetic field $D^2/\eta$ to the timescale for magnetic induction at the flow scale $\ell/U_p$
\cite{Moffatt2019,Tobias2021}:
\begin{linenomath*}
\begin{equation}
	\Rm= \frac{\omega D^2}{\eta} = \Rey_{\ell} \frac{D^2}{\ell^2} \Pm,
	\label{eq:Rm}
\end{equation}
\end{linenomath*}
where $\omega=U_p/\ell$ is the typical vorticity amplitude.
We will use the minimal requirement that $\Rm$ must be greater than unity for dynamo action.
We stress that this requirement is far from sufficient to prove the dynamo capability of the flow, but if fingering convection cannot produce large $\Rm$, it is 
an indication that it will fail to sustain a dynamo.
Note that we consider a system where the scale $D$ of the magnetic field is much larger than the scale $\ell$ of the velocity,
hence there is no requirement for a Reynolds number defined traditionally as $U_p D/\eta$ to be larger than unity, as discussed in 
\citeA{Moffatt2019,Tobias2021}.

In the liquid cores of terrestrial planets, the Prandtl number is estimated to be $\mathcal{O}(10^{-1})$ \cite{Olson2015}. 
Considering the case of relatively weak stratification, $r_{\rho}<\Pran$, for which the flow is the most vigorous, 
we found that $\Rey_{\ell}$ scales approximately as 
$1/\Pran$ or $1/\Pran^{1/2}$. Given that the Prandtl number is not particularly small in planetary cores, both scalings give a similar estimate 
for the local Reynolds number of the fingers, $\Rey_{\ell} =\mathcal{O}(10)$. 
For the azimuthal finger length $\ell$, we found that the theoretical estimate $d/D=(\kappa_t \nu/N_t^2)^{1/4}$ is a good fit for $\ell$ with $\ell\approx 5 d/D$. 
Typical values of $N_t^2$ are generally unknown for planetary cores, so here we will use the values that have been proposed 
for the stable layer at the top of Earth's core (see for instance the summary in \citeA{Gastine2020}). 
Only a rough estimate of $N_t^2$ is needed because of the exponent $-1/4$ in the theoretical scaling. 
Using a mid-range estimate, $N_t^2\approx\Omega^2$ and $\kappa_t\approx 10^{-4}$m$^2/$s,  $\nu\approx 10^{-6}$m$^2/$s, 
we get $\ell\approx \mathcal{O}(1)$m, \ie $D/\ell\approx 10^5$ for a layer thickness of order $100$km. 
Finally, using $\Pm=\mathcal{O}(10^{-6})$, we get $\Rm\approx10^5$. 
Consequently, our estimate of $\Rm$ indicates that fingering convection might be able to produce dynamo action in this system.

\subsection{Axisymmetrisation of the magnetic field}
Next, we assess whether the differential rotation produced by the fingering convection in the stable layer might be 
of sufficient amplitude to axisymmetrise a poloidal magnetic field.
\citeA{Stevenson1982} studied the attenuation of non-axisymmetric magnetic fields passing through a stable layer 
subject to differential rotation by considering the linear interaction of a magnetic field with a radial shear.
This analysis provides a simple approximate relation 
between the attenuation of non-axisymmetric fields through the layer and the shear strength.
To reduce the equatorial dipole such that the dipole tilt becomes smaller than 1$^{\circ}$ at the planetary surface
(as relevant for Saturn \cite{Cao2011} and Mercury \cite{Anderson2012}),
\citeA{Stevenson1982} obtained the condition that $\Rm_0 \geq 30 R_0/D$, where  
the magnetic Reynolds number $\Rm_0$ is based on the radial shear $\omega_0$, $\Rm_0=\omega_0 D^2/\eta$.
Clearly this condition is most easily met for thick stable layers. 
Here we want to determine if the differential rotation obtained in our simulations would meet this requirement. 
Using as a lower bound for the radial shear $\omega_0=U_0/D$ and $D/R_0=0.4$ in our model, 
the zonal Reynolds number $\Rey_0=U_0 D/\eta$ must be at least $75/\Pm$, \ie of the order of $10^8$.
For a Mercury-like core with $\Ek=10^{-12}$, this implies that $\Ro_0$ must be at least $10^{-4}$. 
Figure~\ref{fig:Ro0} shows that this condition is easily met, with $\Ro_0$ reaching this value for 
relatively weak compositional gradients, $\Ra_c \Ek^2\approx 0.01$, or equivalently $|N_c^2|/\Omega^2\approx10^{-3}$.
This simple linear criterion therefore suggests that the amplitude of the differential rotation produced by fingering convection in a thick stable layer 
is sufficiently strong to axisymmetrise the magnetic field. 
However, there are several caveats.
First, the study of \citeA{Stevenson1982} is linear and ignores the effect of the Lorentz forces, so nonlinear calculations are required to verify this result.
Second, the amplitude of the differential rotation is not the only crucial factor in the axisymmetrisation of a magnetic field. 
The direction and equatorial symmetry of the zonal flows are also important as shown by \citeA{Stanley2010}, as certain zonal flow patterns
in the stable layer can disrupt the dynamo in the convective layer underneath, leading to more non-axisymmetric fields. 
Third, our study is limited to a variation of the Ekman number of two decades, hence the extrapolation to much smaller Ekman numbers should be applied with caution.

\section{Conclusion}
\label{sec:conclusion}
This paper presents numerical simulations of double-diffusive convection in a rotating spherical shell that models the stably-stratified layer located at the
top of planetary cores. 
We study the case of a Mercury-like layer, where the temperature gradient is stable ($N_t^2>0$), the compositional gradient is unstable ($N_c^2<0$) and the layer thickness
is 40\% of the outer core radius. 
Fingering convection develops in the form of columnar flows aligned with the rotation axis.
For small compositional and thermal Rayleigh numbers, 
the radial flows have large azimuthal lengths, as reported by previous studies \cite{Monville2019,Mather2021}.
For larger Rayleigh numbers, the radial flows have a sheet-like structure that is elongated in the meridional direction and with small azimuthal lengths. 
The radial variation of the density ratio, $R_{\rho}=|N_t^2|/|N_c^2|$, across the layer leads to some visible changes in the flow; 
notably, the amplitude of the radial flow decays towards the core-mantle boundary as the density ratio 
is the strongest there, and the azimuthal length varies linearly with radius.
We find that the mean azimuthal length follows the scaling law expected for non-rotating planar fingering convection \cite{Stern1960},
varying with the temperature stratification as $|N_t^2|^{-1/4}$. 
For a thermal stratification with $N_t^2\approx\Omega^2$ and for typical core values of the thermal diffusivity and viscosity, 
we obtain a typical azimuthal length of the order of 1m. 
The local Reynolds numbers based on the radial velocity and azimuthal length of the fingers always remain of order $1-10$
for the small Prandtl numbers that are relevant for planetary cores ($\Pran=\mathcal{O}(0.1)$). 
Consequently, fingering convection at large Rayleigh numbers is small scale and laminar.   
This implies small magnetic Reynolds numbers for magnetic fields generated at the flow scale ($\Rm_{\ell}=U_p\ell/\eta\approx10^{-6}$). 
Nevertheless, this does not preclude dynamo action. Indeed the magnetic Reynolds number for magnetic fields generated at the system size (as defined by equation~\eqref{eq:Rm})
can reach large values, 
which might allow to sustain system-size magnetic fields as in models of rotating standard overturning convection \cite{Calkins2015}. 
The dynamo properties of fingering convection can only be fully determined using magneto-hydrodynamical simulations, which we defer to a 
later study.

Zonal flows form across all the parameter range considered here.
The zonal flow is retrograde in the equatorial region outside the tangent cylinder. As the Rayleigh numbers increase and fingering convection onsets
inside the tangent cylinder, the zonal flow becomes prograde in this region. The zonal flow is due to the
latitudinal variations of the axisymmetric composition perturbation. These are produced by the non-uniform transport of composition in the polar and equatorial regions.
We find that this feature is fairly robust to changes of homogeneous boundary conditions
for the temperature or composition fields. 
The typical zonal velocity scales linearly with the compositional Rayleigh number $\Ra_c$,
becoming stronger than the non-axisymmetric velocity at the largest Rayleigh numbers, with  
associated Reynolds numbers of the order of $1000$.
The shear produced by the zonal flows is likely sufficient to axisymmetrise a poloidal magnetic field and reduce the dipole tilt to values smaller than $1^{\circ}$
according to linear considerations \cite{Stevenson1982}.
However, magneto-hydrodynamical simulations are required to check this deduction. 
Indeed, the feedback of the Lorentz forces is known to greatly affect zonal flows produced by convection:
while zonal flows driven by Reynolds stresses tend to be disrupted in dynamo simulations \cite{Aubert2005}, 
zonal flows driven by thermal winds can be amplified in the presence of magnetic fields in magnetoconvection simulations \cite{Mason2022}.
The direction and symmetry of the zonal flow pattern, and the thickness of the stable layer are also known to
affect this process of axisymmetrisation  \cite{Stanley2010,Tian2015}.
Nonetheless, it is interesting that fingering convection has the ability to drive strong zonal flows within the stable layer without
requiring heterogeneous boundary conditions. 

Zonal flows (and associated meridional circulations) are the only large-scale structures observed in our simulations.
The formation of thermo-compositional staircases is not expected to occur in fingering convection at low Prandtl numbers \cite{Traxler2011b}.
However, in numerical simulations of fingering convection in unbounded gradient planar layers, 
\citeA{Brown2013} observed the spontaneous formation of staircases for weak stratifications quantified by the reduced density ratio $r_{\rho}\leq0.003$. 
We performed a small number of simulations at such small values of $r_{\rho}$, but did not observe staircases. 
Although staircases might not form spontaneously in this system, they might survive if started from an initial condition \cite{Moll2017b,Garaud2018}. 
If present, staircases would strongly influence the dynamics of the system, so the forced and spontaneous behaviours of staircases clearly require further investigation.
We did not observe the formation of large-scale vortices as in the planar simulations of rotating fingering convection of \citeA{Sengupta2018}.
In planar simulations in the absence of horizontal anisotropy, the flows in the horizontal plane have no preferred direction,
and so, box-size flows take the form of vortices rather than unidirectional flows (see for instance the case of rotating convection \cite{Guervilly2017c}). 
By contrast, in spherical geometry, the anisotropy introduced by the spherical boundaries leads to a preferred direction for the flows. 
We might therefore expect zonal flows in spherical geometry to be analogous to the large-scale vortices of planar geometry.
However, in our simulations, we showed that the zonal flows are not produced by nonlinear Reynolds stresses. 
The flow is dominated by rotation in most of our simulations (\ie $\Ta^{\ast}$ as defined in \citeA{Moll2017} is smaller than unity),
but it is also always laminar with $\Rey_{\ell}=\mathcal{O}(1)$. By analogy with rotating convection, the nonlinear upscale transfer of 
kinetic energy via the Reynolds stresses might require Reynolds numbers of the order of $100$ \cite{Guervilly2014}. 
This might be achievable in fingering convection at lower Prandtl numbers, as $\Rey_{\ell}$ is inversely proportional to $\Pran$
(equation~\eqref{eq:scaling1} or \eqref{eq:scaling2}). Consequently, this nonlinear process might be more relevant for stellar interiors (where typically $\Pran=\mathcal{O}(10^{-6})$ \cite{Garaud2018}), 
than for planetary interiors.

For stratifications that fall within the range unstable to fingering convection ($1<R_{\rho}<\Le$), we find that 
the minimum background composition gradient required at the onset of fingering convection is very small. Indeed, the minimum 
$\Ra_c$ required for fingering convection is of the order of $100\Ek^{-1}$ (or, equivalently, $|N_c^2|/\Omega^2$ of the order of $10\Ek$) 
in our simulations at $\Ek=\{10^{-5},10^{-4}\}$.
Using the inviscid scaling obtained by \citeA{Monville2019} (\ie the minimum $\Ra_c$ scales as $\Ek^{-1}$)
a minimum background composition gradient of the order $|N_c^2|/\Omega^2\approx 10^{-11}$ at the bottom of the stable layer is sufficient for fingering convection to occur 
in Mercury (using $\Ek\approx10^{-12}$).
As $\Le$ is large (of the order of $10^3$), the range of density ratio unstable to fingering convection is also wide.
Consequently, fingering convection might be a common occurrence in the stable layer of 
planetary cores for which the composition gradient is unstable.
 
Double-diffusive convection in the stable layers of planetary cores remains a relatively unexplored area. 
The interaction of double-diffusive convection with magnetic fields is of considerable interest for planetary magnetism. 
In addition to considering the dynamo properties and the axisymmetrisation of an external poloidal magnetic field, 
the feedback of a magnetic field on double-diffusive convection should also be examined. 
In dynamo simulations in the presence of a stable top layer (without double-diffusive effects), \citeA{Christensen2008} found that a toroidal field 
is stored in the stable layer, so the interaction of the double-diffusive flows with both poloidal and toroidal magnetic fields is relevant.
Many studies of dynamos and magnetoconvection calculations driven by standard overturning convection show the considerable effect that 
magnetic fields can have on convective flows, and notably, the increase of the flow lengthscales \cite{Yadav2016,Aurnou2017}. 
However, magnetic fields likely affect fingering convection very differently, particularly because the fingers size is 
constrained by thermal diffusion.

The interaction with the underlying convective flows is also of great interest. 
Convective overshoot and penetration from the underlying convective layer 
might influence the flows in stable layers, depending notably on the stratification \cite{Dietrich2018,Gastine2020,Bouffard2022}. 
Studies of the dynamics of the stable layer in isolation provides a better understanding of the lengthscales and timescales of this system and should therefore be very useful 
for future coupled double-diffusive models of the whole core.

\section*{Open Research}
The numerical code (XSHELLS) \cite{Schaeffer2013,Schaeffer2017} used for this research is openly available at \url{https://nschaeff.bitbucket.io/xshells/}.
Datasets for this research are available on the Figshare powered Newcastle University research data repository 
(\url{https://data.ncl.ac.uk}) \cite{dataset}. 

\acknowledgments
C.G. acknowledges support from the UK Natural Environment Research Council under grant NE/M017893/1 
and the UK Science and Technology Facilities Council under grant ST/W001039/1.
This research made use of the Rocket High Performance Computing service at Newcastle University.
The author thank the referees for suggestions that have improved the manuscript.


\end{document}